\documentclass[prx,aps,10pt,english,superscriptaddress,twocolumn,amsfonts,amssymb]{revtex4-2}
\usepackage[T1]{fontenc}
\pdfoutput=1
\usepackage{textcomp}
\usepackage{xcolor}
\usepackage{mathrsfs}
\usepackage{amsmath}
\usepackage{bbm} 
\usepackage{amssymb}
\usepackage{graphicx}
\usepackage{esint}
\usepackage{wasysym}
\usepackage[normalem]{ulem}

\makeatletter
\usepackage[english]{babel}
\usepackage{braket}

\usepackage[bookmarks=true,colorlinks,linkcolor=blue,urlcolor=blue,citecolor=blue]{hyperref}

\usepackage{epsfig,graphicx,psfrag,amsmath,amssymb,float}
\usepackage[caption=false]{subfig}

\@ifundefined{showcaptionsetup}{}{%
 \PassOptionsToPackage{caption=false}{subfig}}
\usepackage{orcidlink}
\usepackage{subfig}
\makeatother

\usepackage{babel}

\newcommand{\al}{\alpha}
\newcommand{\be}{\beta}
\newcommand{\g}{\gamma}
\newcommand{\de}{\delta}
\newcommand{\s}{\sigma}
\newcommand{\z}{\zeta}
\newcommand{\lt}{\left}
\newcommand{\rt}{\right}
\newcommand{\lela}{\left\langle}
\newcommand{\rira}{\right\rangle}

\newcommand{\mc}{\mathcal}
\newcommand{\mb}{\mathbf}

\newcommand{\bea}{\begin{eqnarray}}
\newcommand{\eea}{\end{eqnarray}}

\begin{document}
\title{The spin-orbital Kitaev model: from kagome spin ice to classical fractons}
\author{Weslei B. Fontana}
\affiliation{Department of Physics, National Tsing Hua University, Hsinchu 30013, Taiwan}
\affiliation{International Institute of Physics, Universidade Federal do Rio Grande
do Norte, 59078-970 Natal-RN, Brazil}
\author{Fabrizio G. Oliviero}
\affiliation{Center for Theory and Computation, National Tsing Hua University, Hsinchu 30013, Taiwan}
\affiliation{Physics Division, National Center of Theoretical Sciences, Taipei 10617, Taiwan}
\author{Rodrigo G. Pereira}
\affiliation{International Institute of Physics, Universidade Federal do Rio Grande
do Norte, 59078-970 Natal-RN, Brazil}
\affiliation{Departamento de F\'isica, Universidade Federal
do Rio Grande do Norte, 59078-970 Natal-RN, Brazil}
\author{Willian M. H. Natori}
\affiliation{Gleb Wataghin Institute of Physics, University of Campinas, Campinas,
São Paulo 13083-950, Brazil}
\affiliation{Institute Laue-Langevin, BP 156, 41 Avenue des Martyrs, 38042 Grenoble
Cedex 9, France}
\begin{abstract}
We study an exactly solvable spin-orbital model that can be regarded as a classical analogue of the celebrated Kitaev honeycomb
model and describes interactions between Rydberg atoms on the ruby lattice. We leverage its local and nonlocal symmetries to determine the exact partition function and the static structure factor. A mapping between $S=3/2$ models on the honeycomb lattice and kagome spin Hamiltonians
allows us to interpret the thermodynamic properties in terms of a classical kagome spin ice. Partially lifting the symmetries associated with line operators, we obtain a model characterized by   immobile excitations,  called  classical fractons, and a ground state degeneracy that increases exponentially with the length of the system. We formulate a continuum theory that reveals the underlying gauge structure and conserved charges. Extensions of our theory to other lattices
and higher-spin systems are suggested. 
\end{abstract}
\maketitle

\section{Introduction}

The key tasks of condensed matter physics are to describe, classify, and predict emergent phenomena of quantum systems in the thermodynamic limit \citep{MoessnerMoore2021}. An ongoing trend in this branch of physics is the characterization of phases lying outside the Landau-Ginzburg-Wilson paradigm, as is often the case for frustrated magnets \citep{Balents2010,Lacroix2011}.
It has long been hypothesized that these correlated insulators can host quantum spin liquids (QSLs) \citep{Savary2016, Knolle2019}, whose highly entangled ground states and fractionalized excitations are described by effective gauge theories. The first proposals of QSLs were motivated by the effects of geometric frustration, as exemplified by Anderson's resonant valence bond (RVB) state on lattices with triangular plaquettes \citep{Anderson1973}. Later research demonstrated that QSLs turn up as ground states of a class of exactly solvable models displaying exchange frustration, whereby bond-dependent   anisotropic interactions  enhance quantum fluctuations. In this context, the most studied Hamiltonian is the
Kitaev honeycomb model (KHM) \citep{Kitaev2006}, which harbors Majorana
fermion matter excitations coupled to a static $Z_{2}$ gauge field.
Several candidates for QSLs have been synthesized and characterized
over the past decades, raising hopes that this long-sought state will soon 
be found in solid state platforms \citep{NormanRMP2016,Winter2017,ZhouRMP2017,HerrmannsAnRev2018,Takagi_2019,Wang_2023}. 

\begin{figure}
\begin{centering}
\includegraphics[width=.9\columnwidth]{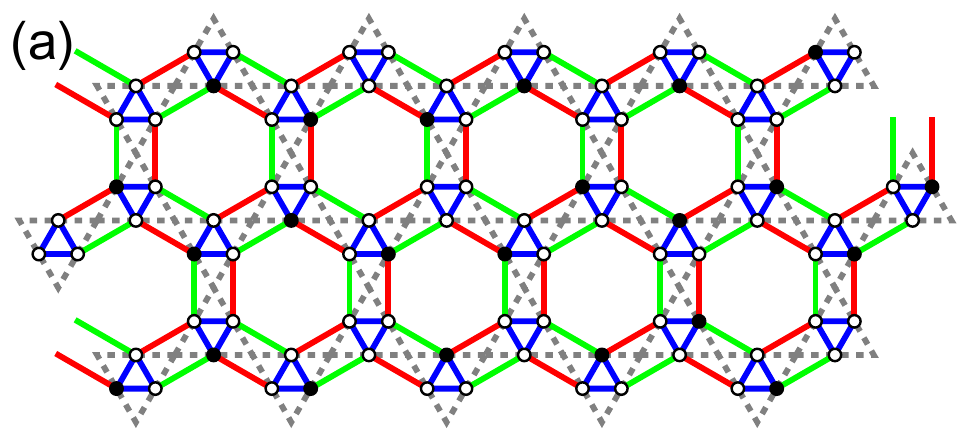}
\par\end{centering}
\begin{centering}
\includegraphics[width=.9\columnwidth]{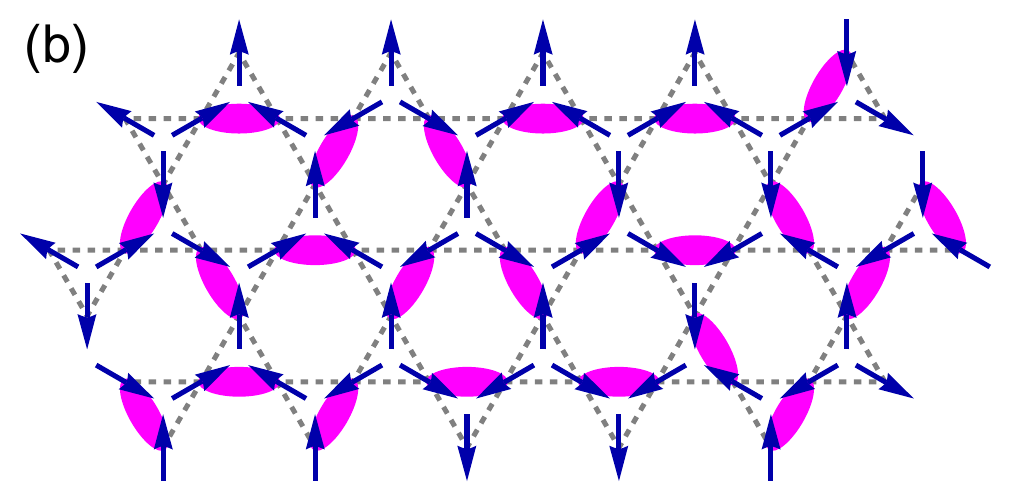}
\par\end{centering}
\caption{\label{fig:1} (a) Rydberg atoms arranged on a ruby lattice. White
(black) dots correspond to atoms in the ground (excited) state. In
this figure, $\langle n_{i}\rangle=1/4$. (b) Any state on the ruby lattice following the specification above can be mapped onto a dimer covering on the kagome lattice. Alternatively, it can also be represented in terms of a spin ice state, as indicated by the arrows. }
\end{figure}

Fractonic spin liquids (FSLs) form a distinct class of phases that can also be found in exactly solvable models and described by exotic quantum  field theories \citep{NandkishoreHermele2019}. FSLs are characterized by a robust subextensive ground state degeneracy and immobile excitations called fractons, which hinder thermal equilibrium at low temperatures \citep{ChamonPRL2005}. In type-I FSLs, fractons can form composite  excitations that are mobile only within a lattice submanifold. The restricted mobility of FSL excitations in real space is reflected on the corresponding field theories, which typically involve tensor gauge fields and dipole conservation laws  \citep{Pretko2017, Pretko2017a, Slagle2017, Slagle2019, Rudelius2021, seiberg2020, Seiberg2021, Burnell2022, Fontana2021, Fontana2022, Fontana2023, Luo2022, Delfino2023, Macedo2024}. 

While fracton order is restricted to exist only in three dimensions \citep{Dominic2020, Haah2021, Haah2021b}, similar physics can emerge in lower dimensions in the form of subsystem protected topological order (SSPT) or spontaneous subsystem symmetry breaking phases (SSSB) \cite{Trithep2018, You2018, Dominic2021, Stephen2019}. Such phases share many similarities with FSLs, such as a geometry-dependent ground state degeneracy and mobility restrictions of the excitations, but differ in the structure of entanglement. In fact, fracton-like behavior has even been identified in classical models constructed from classical spins and Ising variables in two dimensions \cite{HanYan2019,BentonPRL2021,MyersonJain2022,Placke2024} or canonically conjugate variables with dynamics described by Hamilton's equations \cite{Prakash2024}. Here we shall use the term \textit{classical fractons} to refer to these models. 

Quantum simulators based on Rydberg atom arrays \citep{Browaeys2020} provide potential platforms for both QSLs and FSLs. Rydberg atoms trapped on the ruby lattice are already capable of simulating the RVB spin liquid on the kagome lattice \citep{Semeghini2021}. Geometry ensures the isomorphism between excited Rydberg atoms (Fig. \ref{fig:1}) and singlets that occupy kagome lattice bonds \citep{Semeghini2021,Verresen2022}.
The low-energy cold-atom states are univocally mapped onto kagome dimer coverings, which resonate among themselves under tunable transverse fields. The measurements are then compared with the expectation values predicted for a kagome quantum dimer model (KQDM), a projection
of the usual Heisenberg model onto the covering set \citep{ZengPRB1995} that is also exactly solvable \citep{MisguichPRL2002,MisguichPRB2003}.
Alternatively, the KQDM is translated into a honeycomb lattice model of exchange interactions among $S=3/2$ spins subject to longitudinal and transverse fields \citep{Verresen2022}. The zero-field model hosts a dimer liquid corresponding to an equal weight superposition of kagome
dimer coverings. The transverse fields introduce electric, magnetic, or fermionic anyon fluctuations, which drive the dimer liquid to different types of QSLs. In particular, both the spin-1/2 KHM and the kagome RVB occur in this model, thus providing a unified framework for two paradigmatic liquid phases.

In this paper, we study the exchange-frustrated model formulated in Ref. \citep{Verresen2022} in the form of a (classical) spin-orbital Kitaev model (SOKM). This rewriting elucidates both local and nonlocal symmetries, allowing us to uncover the extensive ground state degeneracy associated with the kagome dimer liquid, as well as the exact partition function of the SOKM. Analyzing the nonlocal symmetries, we identify a mechanism to lift the ground state degeneracy leading to a subextensive zero-point entropy. We also discuss the classical fractons and other excitations with restricted mobility that arise in this model.   Finally, we study the continuum limit of this classical fracton model using previously developed techniques for $\Gamma$-matrix models  \citep{Fontana2021, Fontana2022},  illuminating its conservation laws by analogy with Chern-Simons-like theories. 

The paper is organized as follows. Section \ref{sec:The-Classical-Spin-Orbital} presents several exact
results on the SOKM,  connecting the latter with  the well-established physics of frustrated phases on the kagome lattice. In Sec. \ref{subsec:Review-of-Rydberg}, we  review the theoretical models that describe quantum simulations of Rydberg atoms on the ruby lattice in the blockade limit with a
fixed density of excited atoms \citep{Semeghini2021,Verresen2022}. Section \ref{subsec:Exact-Partition-Function}
lists the SOKM symmetries which are then applied to determine its ground state manifold. This analysis automatically leads to classical analogues of Lieb's theorem for gauge fluxes \citep{Lieb1994} and topological order in gapped $Z_{2}$ QSLs \citep{Wen2007}.
The symmetries also determine the level degeneracies, leading to a closed form for the partition function that recovers general features of spin ice thermodynamics \citep{Ramirez1999,Wills2002,MisguichPRB2003,Moller2009}. We use  these results to compute the exact static structure factor in Section \ref{subsection:SSF}, which presents the expected $Z_2$ spin ice character of the ground states. In Sec. \ref{Sec:Classical fractons}, we first introduce a sublattice-dependent rotation that maps the ferromagnetic and antiferromagnetic models and simplifies the notation in the remainder of the paper. We then construct a perturbation that does not commute with local plaquette operators but with strings in two distinct directions. This perturbation lifts the extensive zero-point entropy while also inducing classical fractons.     We show that this model projected on the subspace of SOKM ground states is  similar to the square-lattice fracton model studied in Ref. \citep{HanYan2019}. In Sec. \ref{continummtheory}, we construct an effective field theory for the SOKM. Using this framework, we recover the main features of the lattice model and relate the ground state degeneracy to the algebra of line operators defined in the continuum limit. Finally, we present our conclusions and outlook in Sec. \ref{sec:conclusion}.   The appendix contains explicit expressions for the matrix representations of spin-orbital operators used in this work.

\section{The Classical Spin-Orbital Kitaev Model \label{sec:The-Classical-Spin-Orbital}}

\subsection{Equivalence between the Kagome Dimer Model and the Spin-orbital Kitaev
model \label{subsec:Review-of-Rydberg}}

We briefly review the minimal model for Rydberg atoms arranged on a ruby lattice [see Fig. \ref{fig:1}(a)] as proposed in Ref. \citep{Verresen2022}, both for completeness and to set the notation. The simulations in Ref. \cite{Semeghini2021} were performed using 219 $^{87}\text{Rb}$ ions trapped with optical tweezers, displaying two relevant states at each site: the atomic ground state $ \left| g \rira$ and one Rydberg state $\left| r \rira$. The energy difference between $\left| r \rira$ and $\left| g \rira$ will be called $\hbar \omega_{rg}$. Each atom is modeled by a hard-core boson located at a site $i$ that is described by $n_{i}=0$ or $n_i=1$  if the atom is in the ground or excited Rydberg state, respectively.

An external incoming laser with frequency $\omega_l$ is necessary to populate the Rydberg states and serves to define two additional parameters: the Rabi frequency $\Omega$, and the detuning $\delta$. The Rabi frequency is related with transitions between two states induced by an electromagnetic potential and defined by $\Omega = \lela r \left| \textbf{r} \cdot \textbf{E} \right| g \rira$, in which $\textbf{E}$ is the electric field amplitude of the laser. Repulsive interactions between Rydberg atoms have the form $V_0/r^6$ \cite{Browaeys2020, SaffmanRMP2024}. The competition between the exciting laser field and the Rydberg mutual repulsion defines a blockade radius $R_b = (V_0/\Omega)^{1/6}$, i.e., given a Rydberg state, no other atom can be excited within a radius $r<R_b$. The lattice parameter is chosen to ensure that the six nearest neighbors of a given atom are within $R_b$, thus enforcing a constraint to the Rydberg state arrangements. Finally, the detuning $\delta = \omega_l - \omega_{rg}$ of the incoming laser can favor the occupation of Rydberg states, when positive, and hinder it, when negative.

The classical part of the Hamiltonian describing the aforementioned experimental setup  is \citep{Verresen2022}
\begin{equation}
H_{\text{diag}}=U\sum_{\langle ij\rangle_{\triangle}}n_{i}n_{j}+V\sum_{\langle ij\rangle_{\square}}n_{i}n_{j}-\de\sum_{i}n_{i}.\label{eq:Hdiag}
\end{equation}
Here, $U$ is the Rydberg blockade repulsion  within the lattice triangles and $\langle ij\rangle_{\triangle}$ corresponds to their bonds. The $V$ parameter is the interaction between Rydberg states on sites $i$ and $j$ outside the blockade radius but within nearest-neighbor triangles. Finally, the longitudinal field models the detuning. $H_{\text{diag}}$ is readily converted into an Ising model via $n_{i}= \lt (1 + s_{i}^{z} \rt)/2$. Quantum fluctuations can be introduced by the Rabi frequency $\Omega$ or by including possible hoppings of Rydberg states within elementary triangles, which are modeled by 
\begin{align}
H_{\text{fl}} & =\Omega\sum_{i}s_{i}^{x}+t^{\prime}\sum_{\langle ij\rangle_{\triangle}}\lt(s_{i}^{+}s_{j}^{-}+s_{i}^{-}s_{j}^{+}\rt) \nonumber \\
&+it^{\prime\prime}\sum_{\langle ij\rangle_{\triangle}}\lt(s_{i}^{+}s_{j}^{-}-s_{i}^{-}s_{j}^{+}\rt), \label{eq:Hfl}
\end{align} in which we followed Ref. \cite{Verresen2022} and separate the real and imaginary parts of the hopping. 

\begin{figure}
\begin{centering}
\includegraphics[width=0.5\columnwidth]{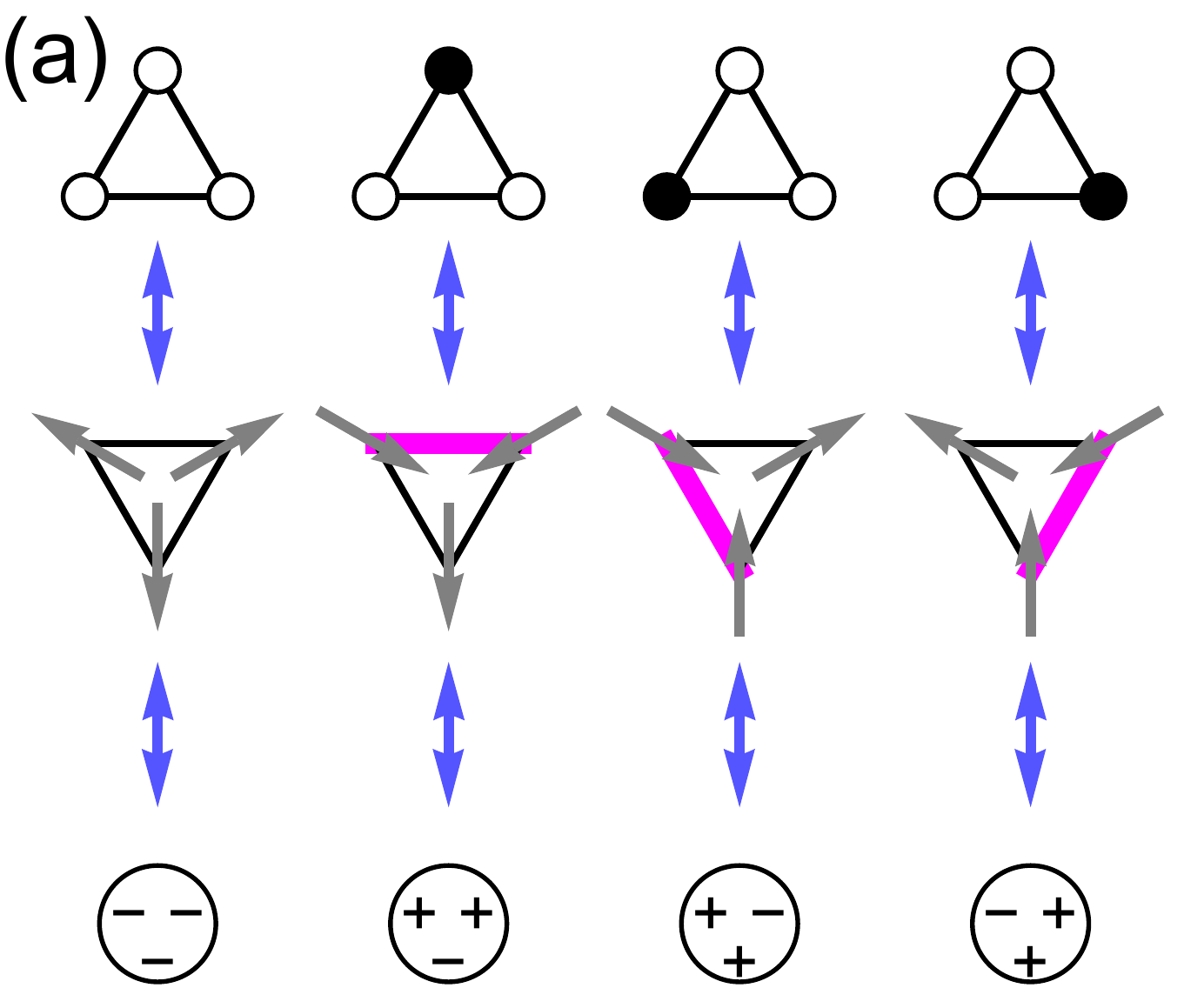}\includegraphics[width=0.5\columnwidth]{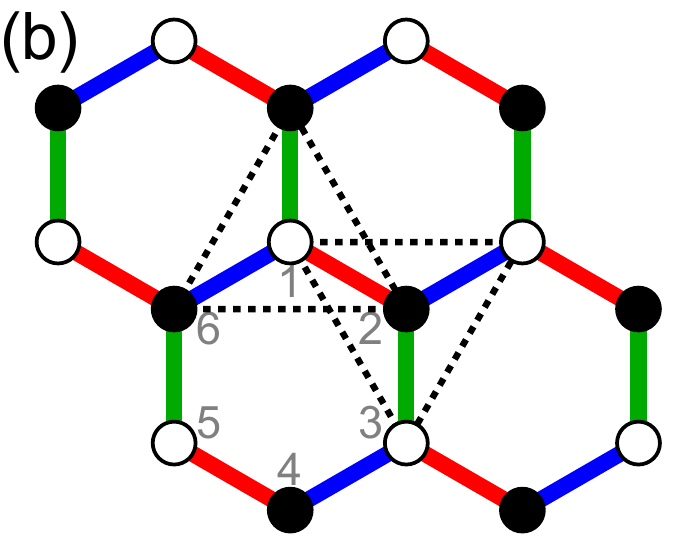}
\par\end{centering}
\caption{\label{fig:equivalences} (a) Equivalence between spin-3/2 states
labeled by $\lt|a^{x},a^{y},a^{z}\rira $ and ruby lattice
states. (b) Section of the honeycomb lattice, in which $x$, $y$, and $z$ bonds are colored blue, red, and green, respectively. Here, black and white dots mark sites on the odd and even sublattices, respectively. A four-site star is formed by the sites on the corners of a dashed triangle plus the site at its center. }
\end{figure}

One ground state of $H_{\text{diag}}$ is illustrated in Fig. \ref{fig:1}(a).Taking the blockade limit  $U\rightarrow\infty$ selects triangles with at most one Rydberg state, while $V$ tends to maximize the distance between
two of these excitations. Since the ruby lattice is the medial lattice of the kagome
\citep{MisguichPRB2003}, states that minimize the interaction energy of $H_{\text{diag}}$ in the sector $\lela n_{i}\rira =1/4$ can be mapped to  kagome dimer coverings; see Fig. \ref{fig:1}(b). It is also possible to set a correspondence between dimer coverings and spin-ice states formed by triangles in two-in-one-out and three-out arrow configurations \citep{ZengPRB1995, MisguichPRL2002, MisguichPRB2003}; see Fig. \ref{fig:equivalences}. Counting arguments show that there are $2^{N+1}$ distinct dimer coverings on the kagome lattice \citep{MisguichPRL2002, MisguichPRB2003}, in which $N$ is the number of unit cells, implying an extensive zero-point entropy. Quantum fluctuations related with the Rabi frequency introduce resonances among the coverings, allowing the simulation of RVB physics \citep{Semeghini2021}. An exhaustive classification of anyon fluctuations was already performed in Ref. \citep{Verresen2022}
and will not be repeated here, where we focus on the classical physics
of $H_{\text{diag}}$. 

The four allowed configurations in each triangle define spin-3/2 states
placed on a honeycomb lattice. The associated observables are more
conveniently written in terms of pseudospin ($\boldsymbol{\s}$) and pseudo-orbitals $(\boldsymbol{\tau})$ variables, each of them satisfying $\z^{x}\z^{y}\z^{z}=i$
($\z=\s,\tau$) and the following algebra \citep{WangPRB2009,Natori2016,Farias2020,HKJinS32_2022,Natori2023}:
\begin{align}
\lt\{ \s_{i}^{\al},\s_{i}^{\be}\rt\}  & =\lt\{ \tau_{i}^{\al},\tau_{i}^{\be}\rt\} =2\de^{\al\be},\nonumber \\
\lt[\s_{i}^{\al},\s_{j}^{\be}\rt] & =2i\de_{ij}\epsilon^{\al\be\g}\s_{i}^{\g},\nonumber \\
\lt[\tau_{i}^{\al},\tau_{j}^{\be}\rt] & =2i\de_{ij}\epsilon^{\al\be\g}\tau_{i}^{\g},\nonumber \\
\lt[\s_{i}^{\al},\tau_{j}^{\be}\rt] & =0.\label{eq:sig_T_relations}
\end{align}
Such operators have been discussed at length in the context of QSLs on Kugel-Khomskii models specially for the SU(4) Heisenberg model \citep{WangPRB2009, Corboz2012, Zhang2021, Jin2023} and exactly solvable extensions of the KHM \citep{Yao2009, Nussinov2009, Yao2011, Carvalho2018, Natori2020, Farias2020, HKJinS32_2022, Natori2023}.
Crucially, the first and last relations in   Eq. (\ref{eq:sig_T_relations})
imply that the operators\begin{equation}
  A_{i}^{\g}=\sigma_{i}^{\g}\tau_{i}^{\g}  
\end{equation}
 commute among themselves. In conjunction with $A_{i}^{x}A_{i}^{y}A_{i}^{z}=-1$,
this implies that they can be diagonalized as follows: 
\begin{align}
A^{x} & =\text{diag}\lt(-1,1,-1,1\rt)=\lt[a^{x}\rt],\nonumber \\
A^{y} & =\text{diag}\lt(-1,1,1,-1\rt)=\lt[a^{y}\rt],\nonumber \\
A^{z} & =\text{diag}\lt(-1,-1,1,1\rt)=\lt[a^{z}\rt].\label{eq:A_rep}
\end{align}
The representation of other spin-3/2 operators in this basis is explicitly provided in Appendix \ref{sec:app}. Throughout this paper, single-site states are labeled by $\lt|a^{x},a^{y},a^{z}\rira $ with $a^{\g}=\pm$ as indicated in Fig. \ref{fig:equivalences}(a). In terms of the spin-ice representation, $a^{\g}=+1$ ($a^{\g}=-1$) is equivalent to an incoming (outgoing) arrow on the triangle. When
depicting the state on a site $i$ of the honeycomb lattice, we place
the $a^{\g}$ closer to the $\g$ bond connected to the site, following Kitaev's convention \citep{Kitaev2006}  shown in  Fig. \ref{fig:equivalences}(b). 

\begin{figure*}
\begin{centering}
\includegraphics[width=0.45\columnwidth]{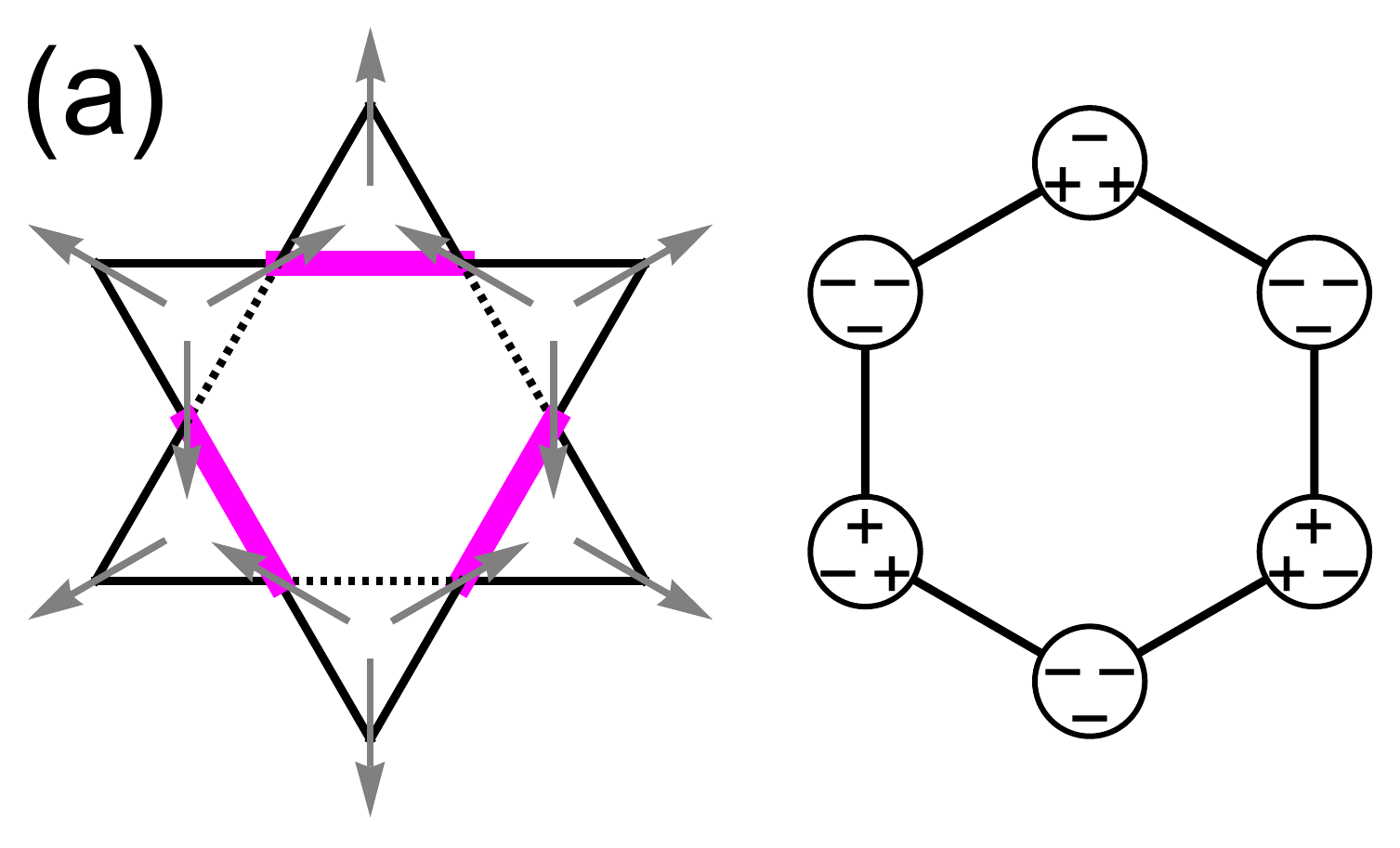}\quad\includegraphics[width=0.45\columnwidth]{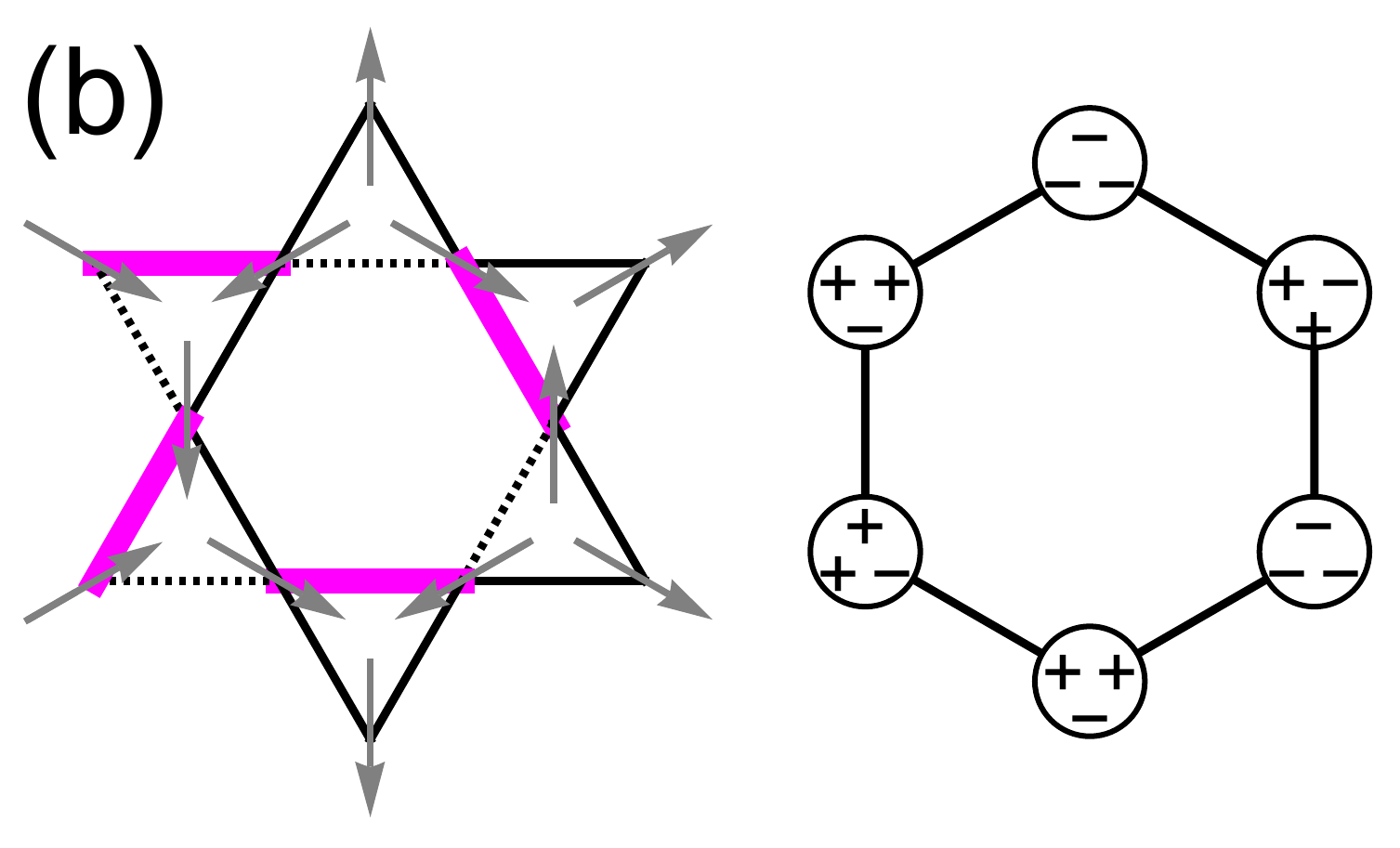}\quad\includegraphics[width=0.45\columnwidth]{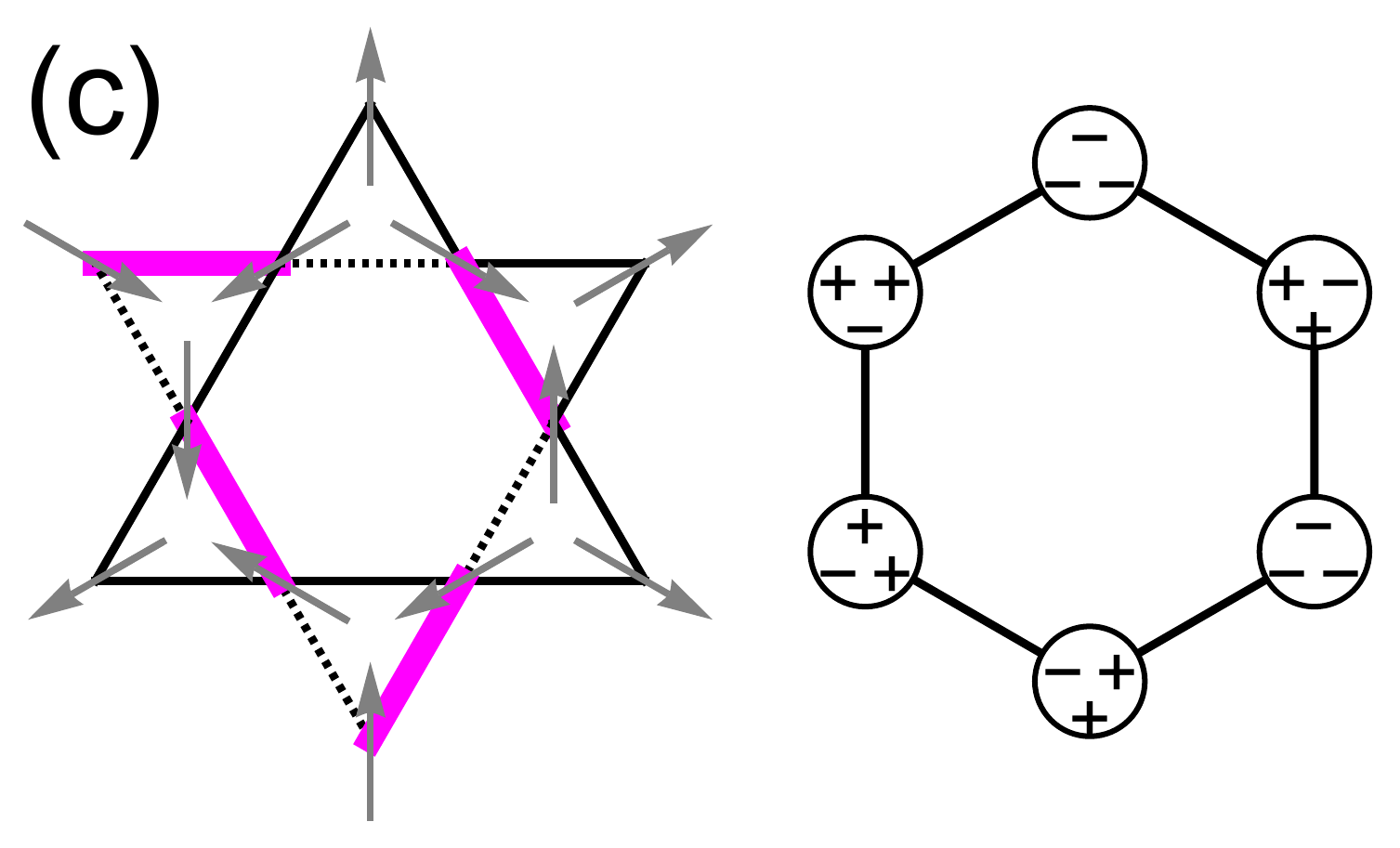}\quad\includegraphics[width=0.45\columnwidth]{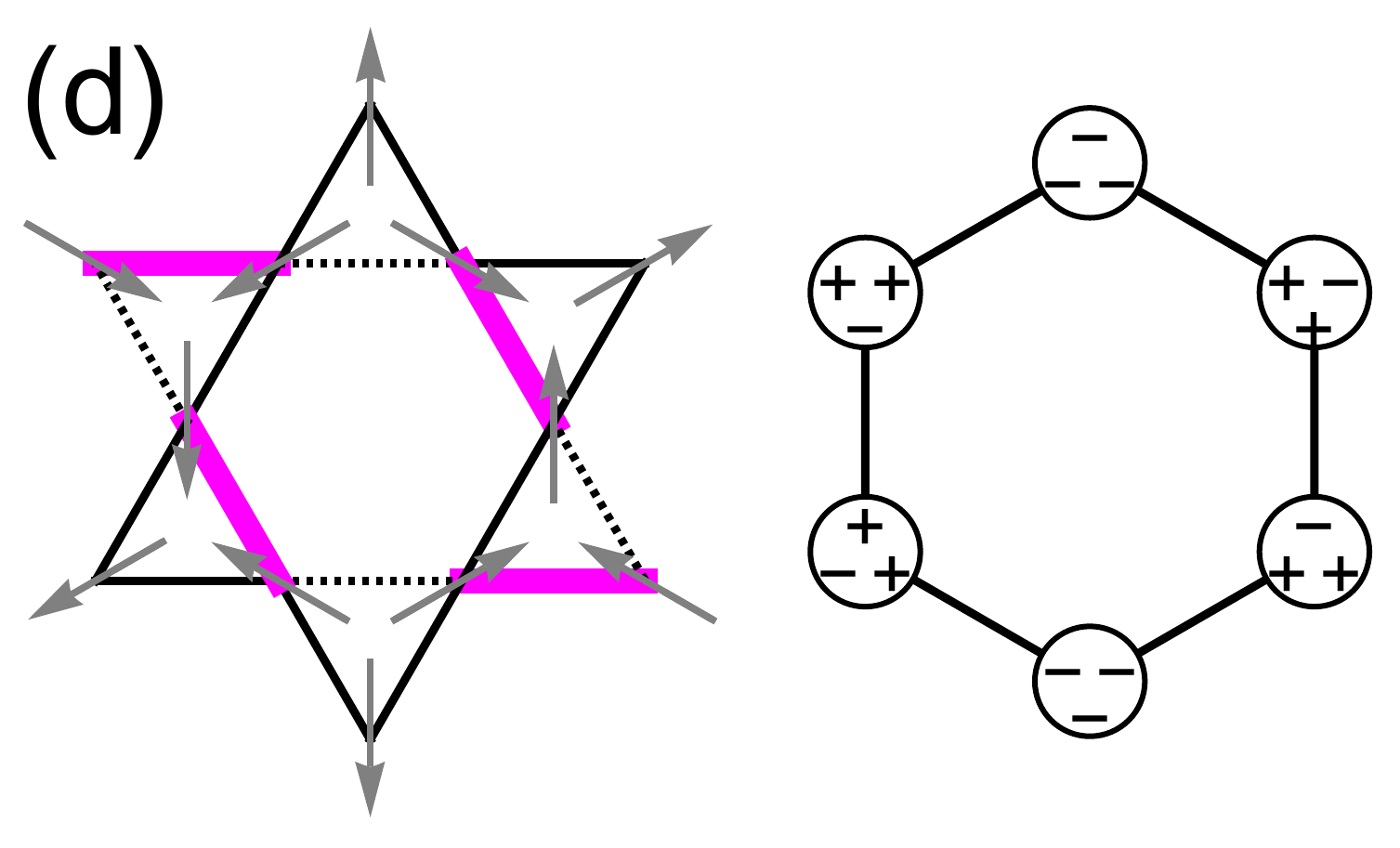}\quad
\par\end{centering}
\begin{centering}
\includegraphics[width=0.45\columnwidth]{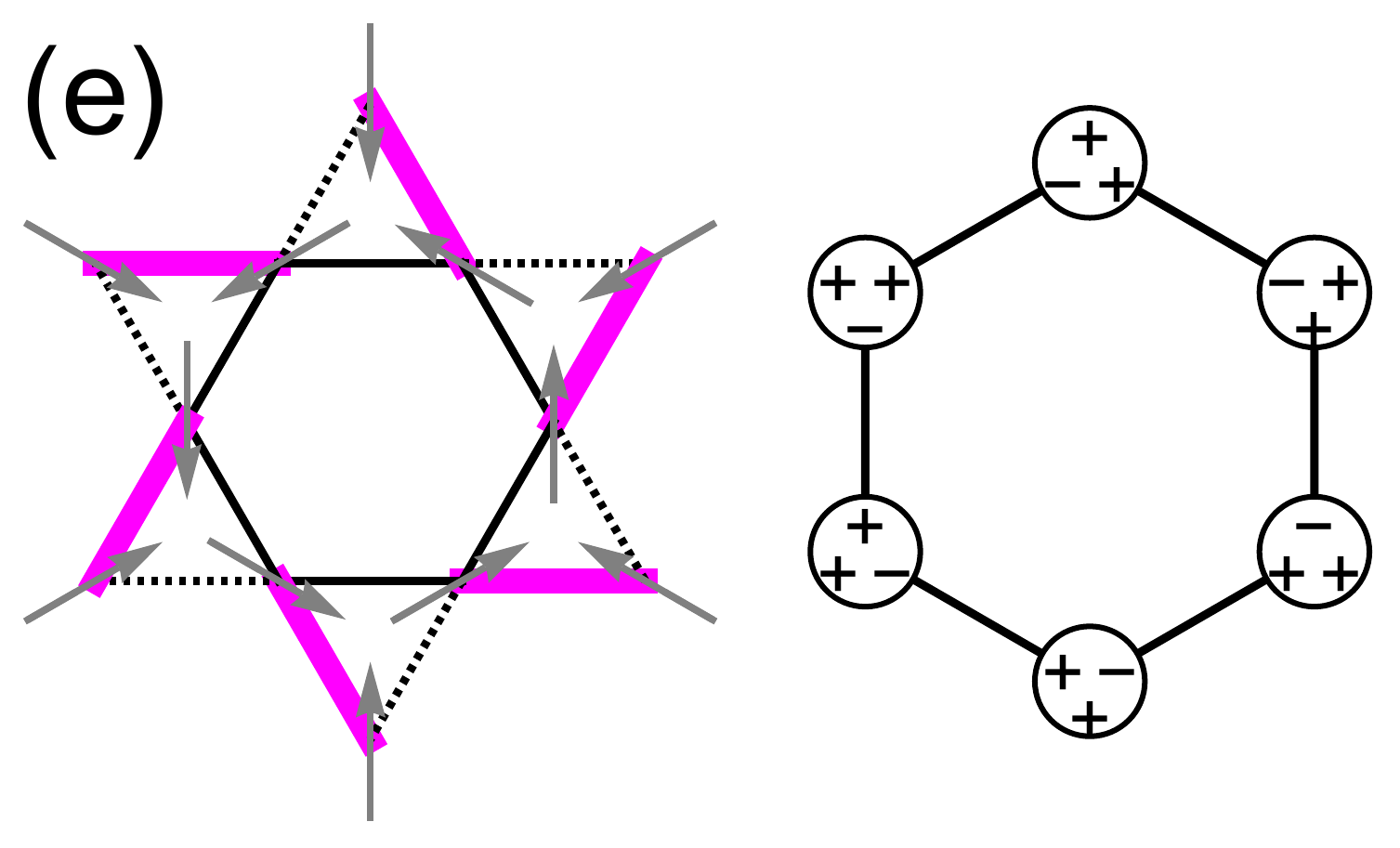}\quad\includegraphics[width=0.45\columnwidth]{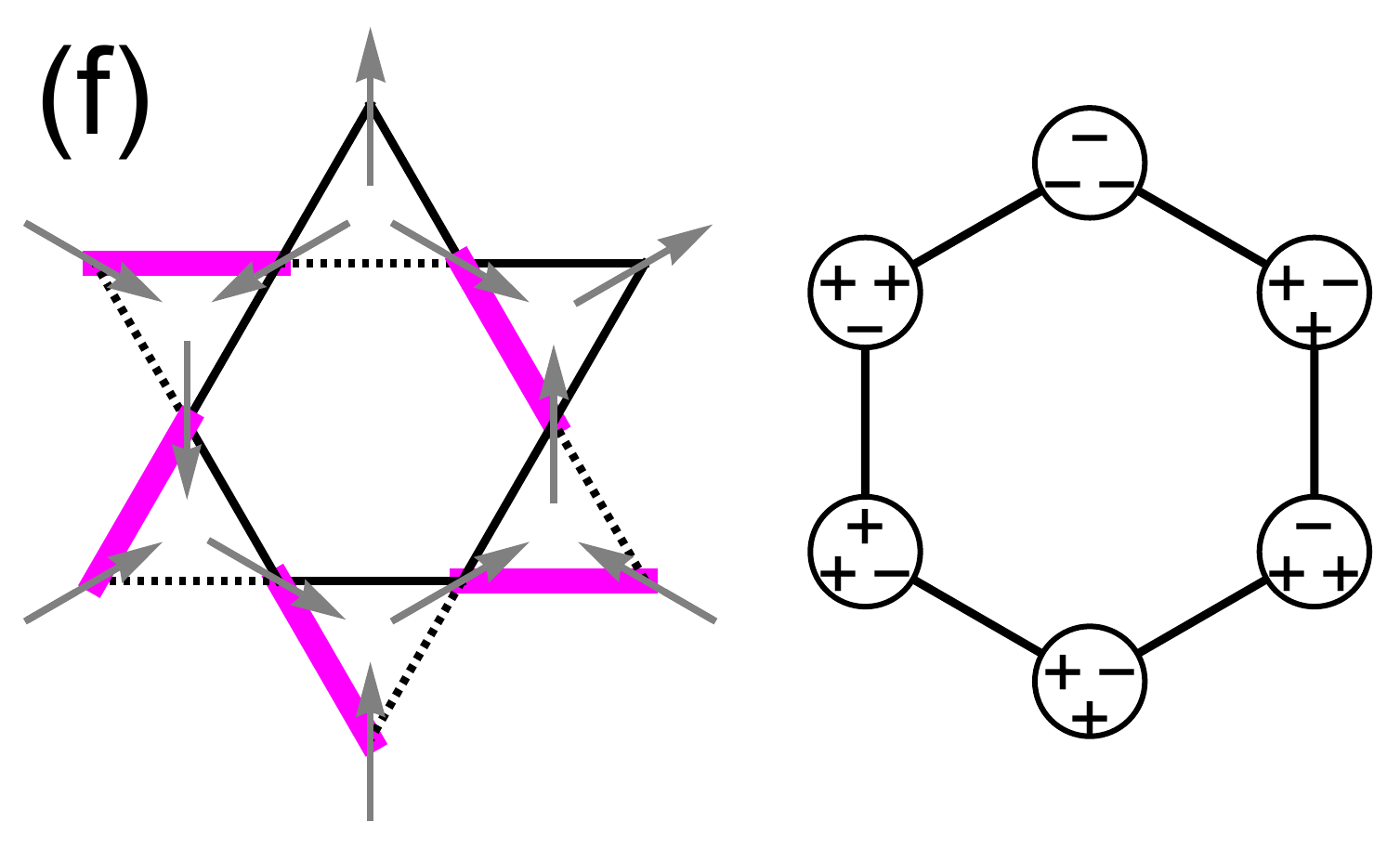}\quad\includegraphics[width=0.45\columnwidth]{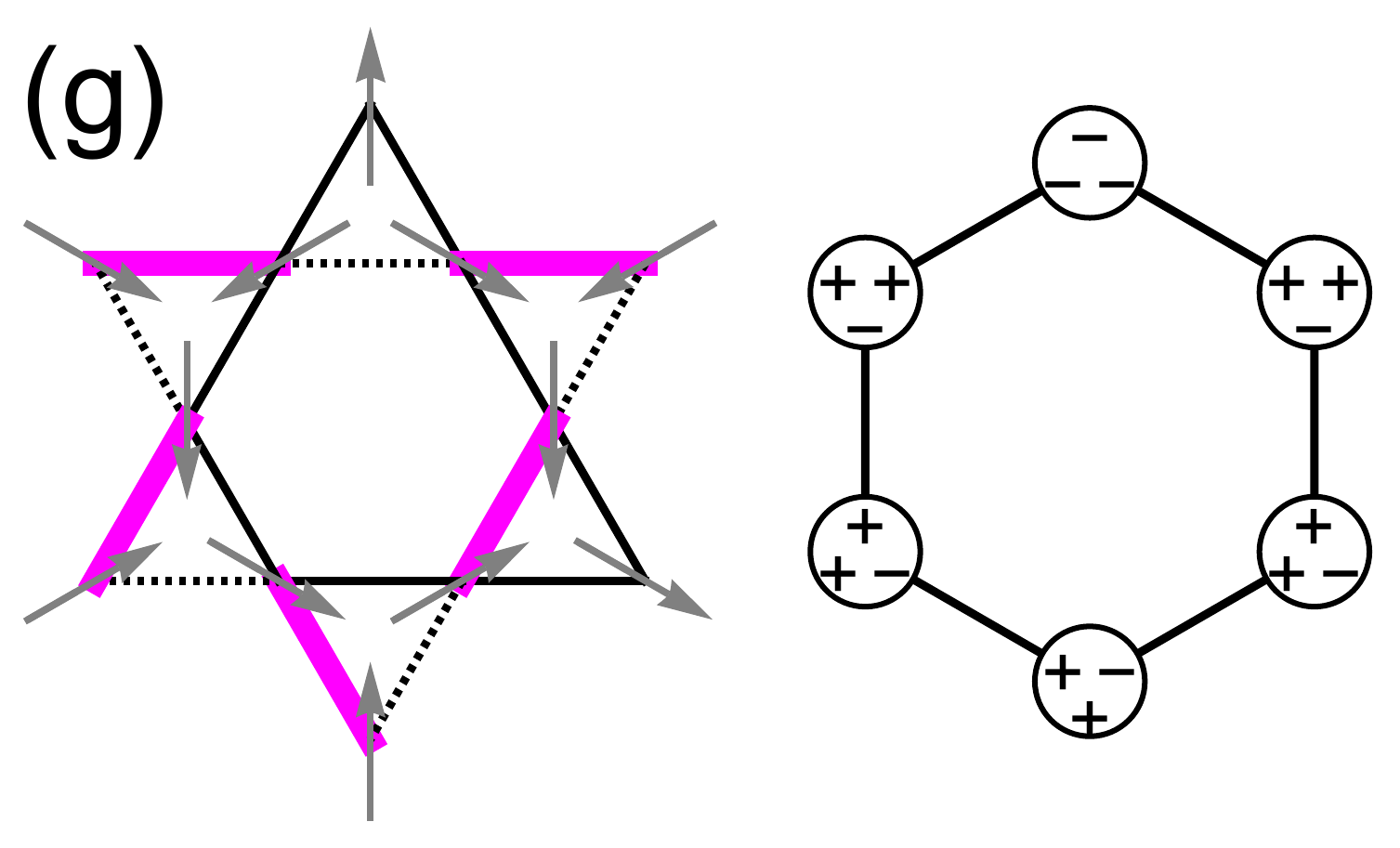}\quad\includegraphics[width=0.45\columnwidth]{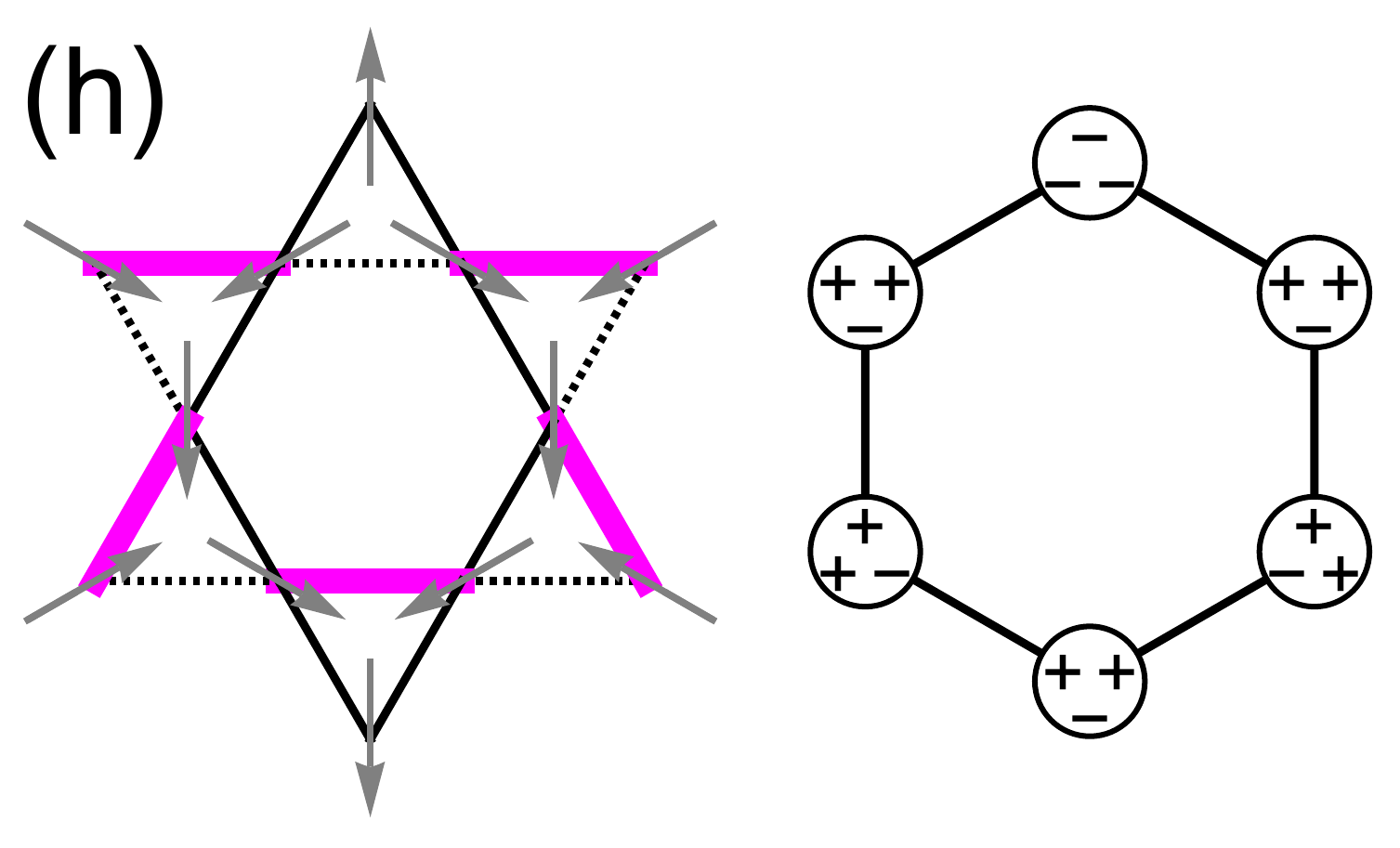}
\par\end{centering}
\caption{\label{fig:loops} Equivalence between loop states on the kagome lattice, as studied in Refs. \citep{ZengPRB1995, MisguichPRL2002, MisguichPRB2003}, and the present paper. All these states clearly satisfy $W^{A}_{p} = +1$, see Eq. (\ref{eq:WpA}), imposing  a constraint on the outward $a^\gamma$ eigenvalues. All previously found loops can be obtained on the honeycomb representation by cyclic permutation around the plaquette and application of the  operator $W_{p}^{\s}$ in Eq. (\ref{Wsigma}).}
\end{figure*}

The Hamiltonian  $H_{\text{diag}}$  in the blockade limit and appropriate excitation density then reads \citep{Verresen2022} 
\begin{align}
H_{\text{diag}} & =H_{\text{SOKM}}+H_{l},\label{HdiagHl}
\end{align}
where $H_{l} = h_{l}\sum_{i} \lt(A_{i}^{x} +A_{i}^{y}+A_{i}^{z}\rt)$ is a longitudinal field with $h_{l}=\lt(V-\de\rt)/4$, which can be set to zero by varying the detuning, and $H_{\text{SOKM}}$ is the SOKM defined as  
\begin{align}
H_{\text{SOKM}} & =J\sum_{\langle ij\rangle_{\g}}A_{i}^{\g}A_{j}^{\g},\label{eq:SOKM}
\end{align}
in which $J=V/4>0$. In view of the bond-dependent anisotropic interactions, this model is a classical analogue of the KHM \citep{Kitaev2006}. Figure \ref{fig:loops} shows the states around an elementary hexagonal plaquette $p$ that minimize the interactions on all bonds. A unifying feature of these states is that the product of all outward $a^{\g}$ equals $+1$. More precisely, the ground state must be in the sector $W_{p}^{A}=+1$ $\forall p$, where
\begin{equation}
W_{p}^{A}=A_{1}^{z}A_{2}^{x}A_{3}^{y}A_{4}^{z}A_{5}^{x}A_{6}^{y},\label{eq:WpA}
\end{equation}
in which the site labels follow Fig. \ref{fig:equivalences}(b). We refer to the subspace spanned by states with $W_{p}^{A}=+1$ $\forall p$ as the zero-flux sector,  establishing a classical analogue of a theorem by Lieb \citep{Lieb1994} that fixes the  ground state sector in the KHM \citep{Kitaev2006}. The $2^{6}/2=32$
possibilities for the outward $\lt\{ a^{\g}\rt\} $ correspond to the 32 loop configurations around hexagons on the kagome lattice \citep{ZengPRB1995}, as illustrated in Fig. \ref{fig:loops}.

\subsection{Symmetries, Exact Spectrum and Partition Function of the Spin-Orbital
Kitaev Model \label{subsec:Exact-Partition-Function}}

\begin{figure}
\begin{centering}
\includegraphics[width=0.45\columnwidth]{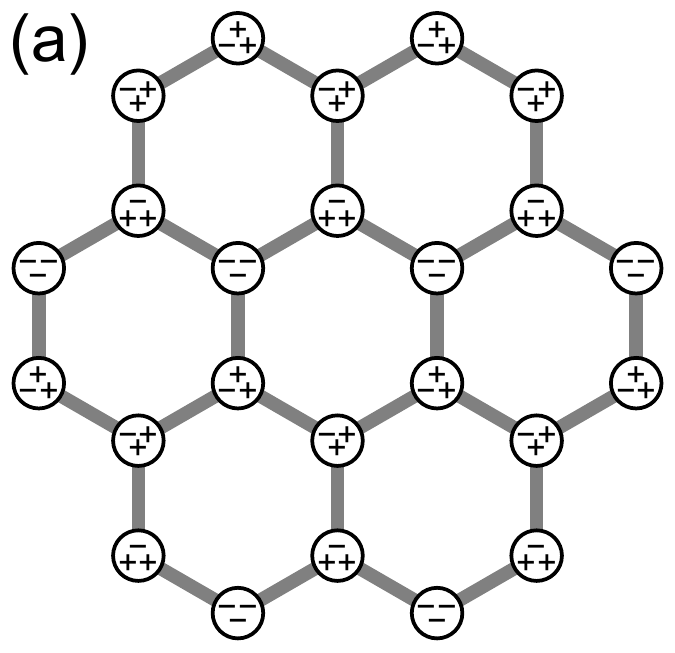}\quad{}\includegraphics[width=0.45\columnwidth]{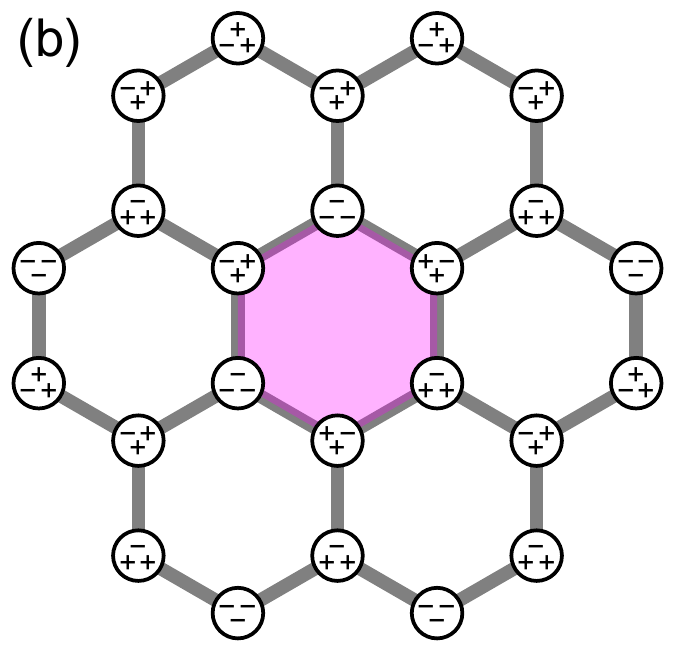}
\par\end{centering}
\begin{centering}
\includegraphics[width=0.45\columnwidth]{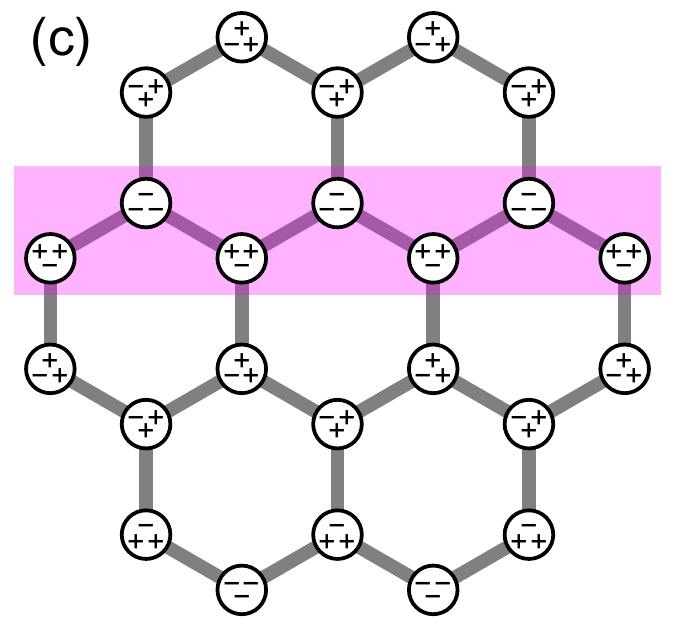}\quad{}\includegraphics[width=0.45\columnwidth]{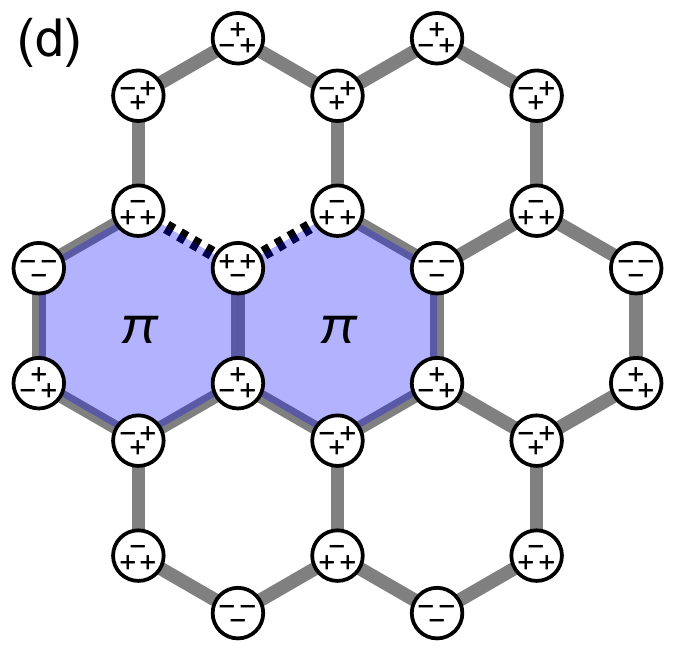}
\par\end{centering}
\caption{\label{fig:AFM_figures} Illustration of the ground state and excitations of the spin-orbital Kitaev model in terms of eigenstates of $A_{i}^{\g}$. (a) Reference ground state $\lt|\phi_{0}\rira $ of the  spin-orbital Kitaev model. (b) Ground state generated by a single plaquette operator, $W_{p}^{\s}\lt|\phi_{0}\rira$. (c)
Ground state generated by a single string operator, $V_{l_{xy}}^{\s}\lt|\phi_{0}\rira$. (d) Single-flip excitation leading to two unsatisfied bonds and two $\pi$-flux plaquettes.}
\end{figure}

The main properties of the SOKM  follow from its local symmetries. Assuming periodic boundary conditions, we can identify the ground state manifold starting from the reference ground
state $\lt|\phi_{0}\rira $ shown in Fig. \ref{fig:AFM_figures}(a). In analogy to Eq. (\ref{eq:WpA}), we define the plaquette operators\begin{equation}
    W_{p}^{\s}=\s_{1}^{z}\s_{2}^{x}\s_{3}^{y}\s_{4}^{z}\s_{5}^{x}\s_{6}^{y}.\label{Wsigma}
\end{equation}
Such plaquette operators still commute with Eq. (\ref{eq:SOKM}) while locally changing $\lt|a^{x},a^{y},a^{z}\rira$ of sites pertaining to $p$. This observation allows the construction of the following ground state family: 
\begin{equation}
\lt|\psi_{0}\lt(\lt\{ S_{p}\rt\} \rt)\rira =\prod_{p=1}^{N}\lt(W_{p}^{\s}\rt)^{S_{p}}\lt|\phi_{0}\rira ,\label{eq:Wp_states}
\end{equation}
in which $S_{p}=0,1$; see Fig. \ref{fig:AFM_figures}(b) for a minimal
example. We could also define plaquette operators $W_p^\tau$ using the pseudo-orbital operators, but the relation $W_p^\sigma W_p^\tau=W_p^A$  implies that the states generated by applying $W_p^{\tau}$ differ from Eq. (\ref{eq:Wp_states}) only by a global phase factor, thus being physically equivalent. 
To count the number of states defined by Eq. (\ref{eq:Wp_states}), let us recall that the $W_{p}^{\s}$ operators are not independent in periodic boundary conditions, but must satisfy $\prod_p W_{p}^{\s} = i^{N_{\text{sites}}}$ due to the identity $\s^{x}\s^{y}\s^{z}=i$. We take the number of plaquettes $N$ to be even, so that the product of these operators is set to $+1$. The $N-1$ independent plaquettes then define $2^{N-1}$ distinct ground states.

In addition to the local symmetries, the SOKM also commutes with nonlocal
operators defined over noncontractible strings. Consider, for example, \begin{equation}
 V_{l_{xy}}^{\s}=\prod_{i\in l_{xy}}\sigma_{i}^{z},   
\end{equation}
which flips $\lt\{ a_{i}^{x}\rt\} $ and $\lt\{ a_{i}^{y}\rt\} $
along an $xy$ zigzag chain labeled by $l_{xy}$; see Fig. \ref{fig:AFM_figures}(c). A new ground state $\lt|\Phi_{0}\rira $
can be written as follows:
\begin{equation}
\lt|\Phi_{0}\lt(\lt\{ S_{l_{xy}}\rt\} \rt)\rira =\prod_{l_{xy}=1}^{L_{xy}}\lt(V_{l_{xy}}^{\s}\rt)^{S_{l_{xy}}}\lt|\phi_{0}\rira ,\label{eq:str_gs}
\end{equation}
in which $S_{l_{xy}}=0,1$ and $L_{xy}$ is the number of $xy$ chains.
An even number of $V_{l_{xy}}^{\s}$ corresponds to the application
of $W_{p}^{\s}$ operators around the honeycomb lattice following
a set of $xy$ chains, but an odd number of $V_{l_{xy}}^{\s}$
cannot be generated using this operation. This even-odd distinction
classifies the ground states by the
parity of the string operators connecting it to the reference ground
state $\lt|\phi_{0}\rira $. Since the lattice is two dimensional,
we can repeat this discussion for another set of noncontractible loops, say, the ones along $yz$ chains labeled by  $l_{yz}=1,\dots,L_{yz}$. As a consequence, the parity of the number of noncontractible strings on a torus defines $2^{2}=4$ ground state sectors, in each of which we can generate the states in Eq. (\ref{eq:Wp_states}). The resulting ground state degeneracy is $4\times 2^{N-1}=2^{N+1}$, for $N=L_{xy}L_{yz}$  unit cells. Clearly,  this classical analogue of $Z_{2}$  topological degeneracy \citep{Wen2007}
is essential to recover the aforementioned degeneracy related to kagome dimer coverings \citep{MisguichPRL2002,MisguichPRB2003}. Note also that all states in the ground state manifold obey the relation $A_i^\gamma+A_j^\gamma=0$ for every bond $\langle ij\rangle_\gamma$. Summing over all bonds, we obtain   $\sum_{i,\gamma} A_i^\gamma=0$. As a consequence, the  longitudinal field term $H_l$ in Eq. (\ref{HdiagHl}) has vanishing projection in the ground state sector. Therefore, the ground state degeneracy is robust against  small changes in the detuning that  generate a weak longitudinal field, $|h_l|\ll J$. Only for $0<h_l\sim J$ do we expect a transition to a polarized phase where $a_i^\gamma=-1$ $\forall i,\gamma$. 

We can also use the symmetries to calculate the partition function of the SOKM. Figure \ref{fig:AFM_figures}(d) represents
a minimally excited state of energy $4J$, obtained by applying a local operator $\sigma_{i}^{\g}$ on a ground state. This operator preserves the eigenvalue $a_{i}^{\g}$ while flipping the other two eigenvalues, $a_{i}^{\al},a_{i}^{\be}$, thus changing the interaction energy on the $\alpha,\beta$ bonds. This excited state contains two visons  \citep{MisguichPRL2002,MisguichPRB2003} associated with $W_p^A=-1$, equivalent to  a $\pi$ flux through the plaquette. Note that visons are always created or annihilated in pairs. Hence, the spectrum reads 
\begin{equation}
E_{n}=-3NJ+4nJ,\qquad n\geq 0.\label{eq:En}
\end{equation} 
Visons can be separated by arbitrary distances without an energy cost by applying $\s_j^\gamma$ along open strings. Combining the distinct ways of distributing the $2n$ unsatisfied
bonds with the degeneracy arising from the symmetry operators, we obtain the degeneracy of each energy level:
\begin{equation}
g\lt(E_{n}\rt)=2^{N+1}\lt(\begin{array}{c}
3N\\
2n
\end{array}\rt),\label{eq:deg_En}
\end{equation}
whose summation recovers the Hilbert space dimension. Using Eqs.
(\ref{eq:En}) and (\ref{eq:deg_En}), we obtain   the partition function 
\begin{eqnarray}
Z & =&\sum_{n}g\lt(E_{n}\rt)e^{-\be E_{n}}\nonumber \\
 & =&2^{N}e^{3NJ\be}\lt[\lt(1+e^{-2\be J}\rt)^{3N}+\lt(1-e^{-2\be J}\rt)^{3N}\rt].\label{eq:Z}
\end{eqnarray}

The partition function allows us to compute the thermodynamics of the SOKM,
which displays the qualitative behavior expected for spin ice \citep{Ramirez1999,Wills2002,MisguichPRB2003,Moller2009}.
The  free energy $F=-\frac{1}{\be}\ln Z$ has the usual  high-temperature  limit  given by 
\begin{align}
F\lt(T\rightarrow\infty\rt) & =-3NJ-2Nk_{B}T\ln4,
\end{align}
which recovers the expected entropy per site of $s_{\infty}=k_{B}\ln4$.
On the other hand, the low-temperature limit $T\rightarrow0$ yields
\begin{align}
F\lt(T\rightarrow0\rt) & =-3NJ-\lt(N+1\rt)k_{B}T\ln2,
\end{align}
which reflects the $2^{N+1}$-fold ground state degeneracy and leads
to an entropy per site of $s_{0}=\frac{1}{2}k_{B}\ln2$ in the thermodynamic limit. Such residual entropy is a defining characteristic of spin ice systems and can be experimentally estimated using specific heat integration \citep{Ramirez1999, Wills2002, MisguichPRB2003, Moller2009}.
From the exact partition function, we can also calculate the specific heat and entropy per site at arbitrary temperatures. The result is shown in Fig. \ref{fig:observables}(a), where the dashed lines highlight the low- and high-temperature limits. Notice that the SOKM does not exhibit a phase transition but rather a crossover connecting the spin-ice and the paramagnetic regimes.

We end this section by comparing our work with other classical Kitaev-like models in which the thermodynamics is also exactly computed \cite{Mizoguchi2022, Mizoguchi2024}. These models are defined on the decorated honeycomb lattice and rely on the conservation of certain components of the spin operators that live on the bonds of the honeycomb lattice. This allows one to map the interacting problem onto a set of noninteracting spins under position-dependent magnetic fields. The partition functions in these models are then given by an exponentially large degeneracy factor multiplied by single-particle partition functions. This should be contrasted with our model, where the conserved quantities are plaquette and closed-loop operators that keep the interactions always active. The partition function in Eq. (\ref{eq:Z}) is then different from those obtained in mean-field theories, being better understood in terms of a spin ice phase.

\subsection{Static Structure Factor}\label{subsection:SSF}

\begin{figure}
\begin{centering}
\includegraphics[width=0.6\columnwidth]{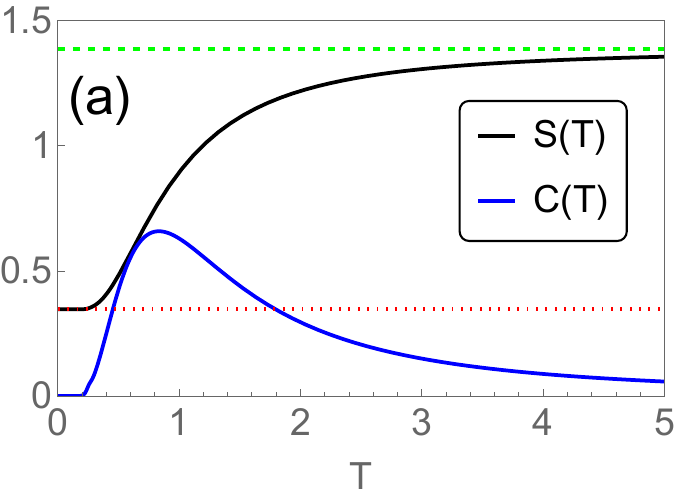}\includegraphics[width=0.4\columnwidth]{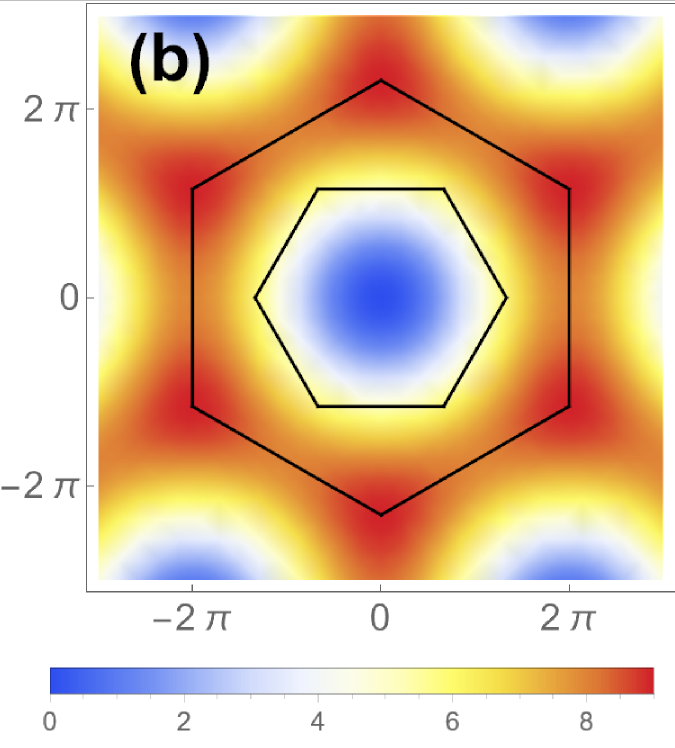}
\par\end{centering}
\caption{\label{fig:observables} (a) Entropy $S(T)$ and specific heat $C(T)$ per site of the spin-orbital Kitaev model (in units of $k_{B}$) as a function of  temperature in units of $J/k_{B}$. The dotted line indicates the zero-point entropy of $\frac{1}{2}\ln2$, while the dashed line corresponds to the high-temperature
entropy of $\ln4$. (b) Static structure factor of the spin-orbital Kitaev model. The smaller hexagon corresponds to the boundaries of the first Brillouin zone, while the larger plaquette connects neighboring spectral weight maxima. }
\end{figure}

We now turn to the calculation of the static structure factor
\begin{equation}
S\left(\mathbf{q},T\right)=\frac{1}{N}\text{Tr}\left[\hat{\rho}\left(T\right)\sum_{\alpha,\beta}\sum_{i,j}A_{i}^{\alpha}A_{j}^{\beta}e^{i\mathbf{q}\cdot\left(\mathbf{r}_{i}-\mathbf{r}_{j}\right)}\right],
\end{equation}
in which $\hat{\rho}(T)= Z^{-1} \sum_\psi \exp(-\beta E_\psi) \left| \psi \right\rangle \left\langle \psi \right|$ is the thermal density matrix defined over all eigenstates $ \left| \psi \right\rangle$. We are particularly
interested in the low-temperature limit, where the correlation function
simplifies to 
\begin{equation}
\lim_{T\rightarrow0}S\left(\mathbf{q},T\right)=\sum_{\underset{i,j}{\psi_{0},\alpha,\beta,}}\frac{\left\langle \psi_{0}\left|A_{i}^{\alpha}A_{j}^{\beta}\right|\psi_{0}\right\rangle }{2^{N+1}N}e^{i\mathbf{q}\cdot\left(\mathbf{r}_{i}-\mathbf{r}_{j}\right)},
\end{equation}
in which $\psi_{0}$ runs over the ground state manifold. It is then convenient to define 
\begin{equation}
\left|0\right\rangle \equiv \frac{1}{\sqrt{2^{N+1}}}\sum_{\psi_{0}}\left|\psi_{0}\right\rangle .
\end{equation}
Since 
\begin{align}
A_{i}^{\alpha}A_{j}^{\beta}\left|\psi_{0}\right\rangle  & =a_{i}^{\alpha}a_{j}^{\beta}\left|\psi_{0}\right\rangle ,\,a_{i}^{\alpha},a_{j}^{\beta}=\pm1,\nonumber \\
\left\langle \psi_{0}^{\prime}|\psi_{0}\right\rangle  & =\delta_{\psi_{0}^{\prime},\psi_{0}},\label{eq:psi0_prop}
\end{align}
we can write the $T\rightarrow0$ correlation as 
\begin{equation}
S\left(\mathbf{q}\right)=\frac{1}{N}\sum_{\underset{i,j}{\alpha,\beta,}}\left\langle 0\left|A_{i}^{\alpha}A_{j}^{\beta}\right|0\right\rangle e^{i\mathbf{q}\cdot\left(\mathbf{r}_{i}-\mathbf{r}_{j}\right)}.\label{eq:static}
\end{equation}

The $\left|0\right\rangle $ state is the honeycomb representation of the kagome dimer liquid discussed in Refs. \citep{MisguichPRL2002,MisguichPRB2003}, which allows  us to assert the ultrashort correlation functions
\begin{equation}
\left\langle 0\left|A_{i}^{\alpha}A_{j}^{\beta}\right|0\right\rangle =\delta^{\alpha\beta}\times\begin{cases}
1, & \text{if }i=j,\\
-1, & \text{if }i,j\in\left\langle ij\right\rangle _{\alpha}\\
0, & \text{otherwise}.
\end{cases}\label{eq:NN_corr}
\end{equation}
The first two equations are explained by the antiferromagnetic Ising
couplings of $Z_{2}$ local states. On the kagome lattice, the last equation reflects the independence of dimer states for triangles that do not share a vertex. This is the classical analogue of the KHM correlation function, which is known to vanish beyond nearest neighbors from analytical arguments \citep{Baskaran2007,Baskaran2008}. Substituting Eq. (\ref{eq:NN_corr}) in 
 Eq. (\ref{eq:static}), we find
\begin{align}
S\left(\mathbf{q}\right) & 
=4\sum_{\gamma}\sin^{2}\left(\frac{\mathbf{q}\cdot\boldsymbol{\delta}_{\gamma}}{2}\right),
\end{align}
where $\boldsymbol{\delta}_{z}=\frac{\sqrt{3}}{3}\left(0,-1\right)$,
$\boldsymbol{\delta}_{x,y}=\frac{\sqrt{3}}{3}\left(\pm\frac{\sqrt{3}}{2},\frac{1}{2}\right)$ are the nearest-neighbor vectors of the honeycomb lattice.

The structure factor $S\left(\mathbf{q}\right)$ plotted in Fig.
\ref{fig:observables}(b) displays no signature of conventional long-range order,
as expected from a classical spin liquid. The momentum dependence is commensurate with the lattice periodicity and consistent with the $Z_3$ rotational invariance of the model. The correlations
are weak in the neighborhood of the $\Gamma$ point, remaining relatively
small for all points inside the first Brillouin zone. The spectral
weight is maximal at some reciprocal lattice points and remains large
along the edges connecting them. Such a pattern is reminiscent of
the spin ice structure factor, with the key difference that there
are no pinch point singularities in the SOKM due to the ultrashort
correlations.

\section{Classical Fractons \label{Sec:Classical fractons}}

\begin{figure}
\begin{centering}
\includegraphics[width=1\columnwidth]{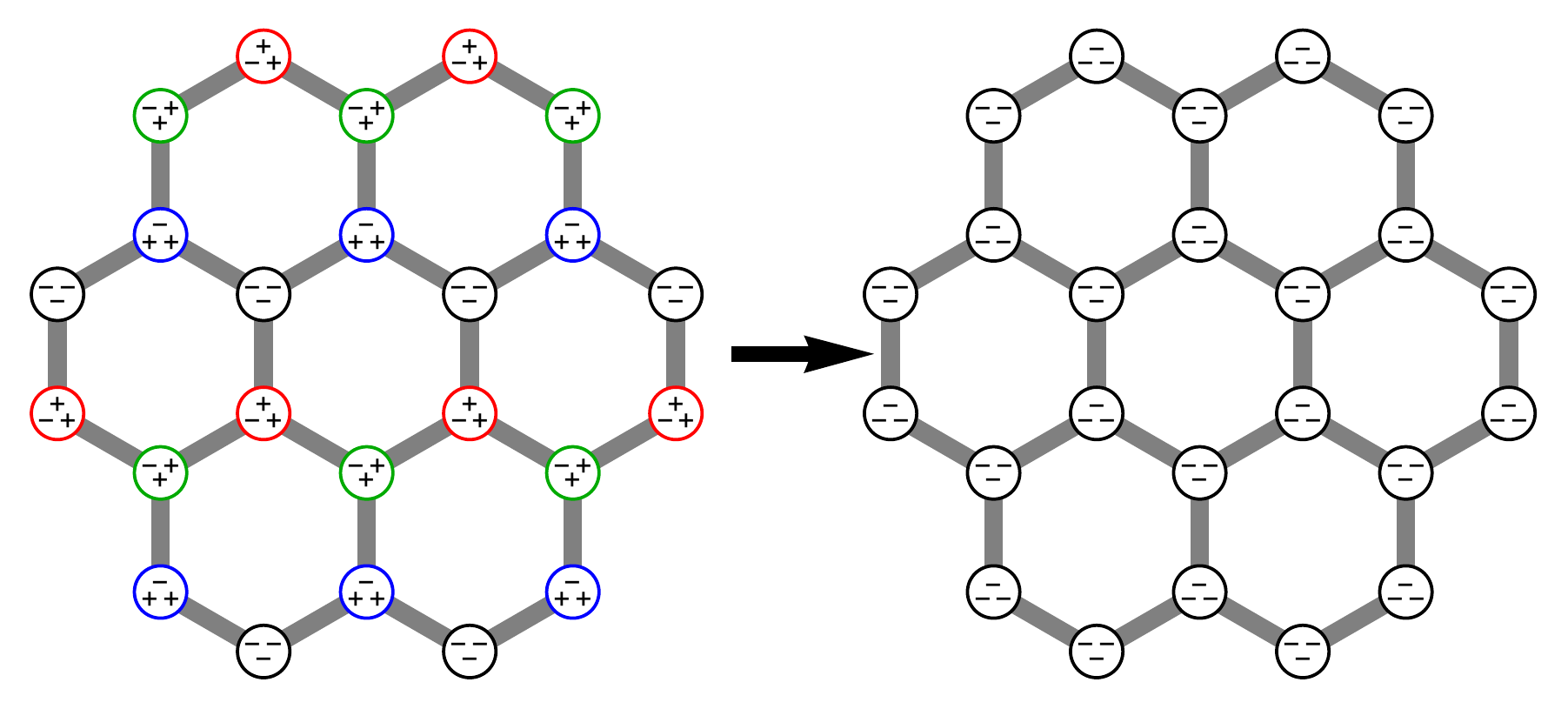}
\par\end{centering}
\caption{\label{fig:Klein_like} Illustration of the Klein-like transformation   applied to the reference ground state. We   perform a $\pi$ rotation around the $\s^{z}$ axis on blue sites, $\s^{x}$ on red sites, and $\s^{y}$ on green sites. This transformation effectively invert the sign of the  exchange coupling, leading to the ferromagnetic model in Eq. (\ref{eq:SOKM_2}).}
\end{figure}

We are now ready to study how to lift the local symmetries associated with $W_{p}^{\s}$ while retaining string symmetries $V_{l}^{\s}$ in a manner that induces fracton excitations. To simplify the onward discussion, we will work on a rotated frame characterized by a site-dependent pseudospin rotation 
$A_{j}^{\g}\rightarrow e^{i\pi\sigma_{j}^{\al}/2}A_{j}^{\g}e^{-i\pi\sigma_{j}^{\al}/2}$, with the rotation axis  chosen locally 
as indicated in Fig. \ref{fig:Klein_like}. In this frame, the reference ground state $\lt|\phi_{0}\rira $ is transformed into a
``ferromagnetic'' reference state based on $\lt| a^{x},a^{y},a^{z}\rira =\lt| -,-,-\rira $. 
More generally, this sublattice-dependent rotation is akin to a Klein transformation in the Kitaev-Heisenberg model \citep{Kimchi2014,ChaloupkaPRB2015,Natori2018}
that effectively inverts the sign of the exchange coupling: 
\begin{equation}
H_{\text{SOKM}}=-J\sum_{\lela ij\rira _{\g}}A_{i}^{\g}A_{j}^{\g}.\label{eq:SOKM_2}
\end{equation}
We stress that the symmetries detailed in the previous section remain valid for the ferromagnetic model. Within this description, the next figures can be unburdened from many sign factors by no longer displaying the $\lt| -,-,-\rira $ state. We will also denote the other local states by \begin{eqnarray}
\left|x\right\rangle&\equiv \lt|-,+,+\rira,\nonumber\\
\left|y\right\rangle&\equiv \lt| +,-,+\rira,\nonumber\\
\left|z\right\rangle&\equiv \lt| +,+,-\rira.
\end{eqnarray}

\begin{figure*}
\begin{centering}
\includegraphics[width=1\textwidth]{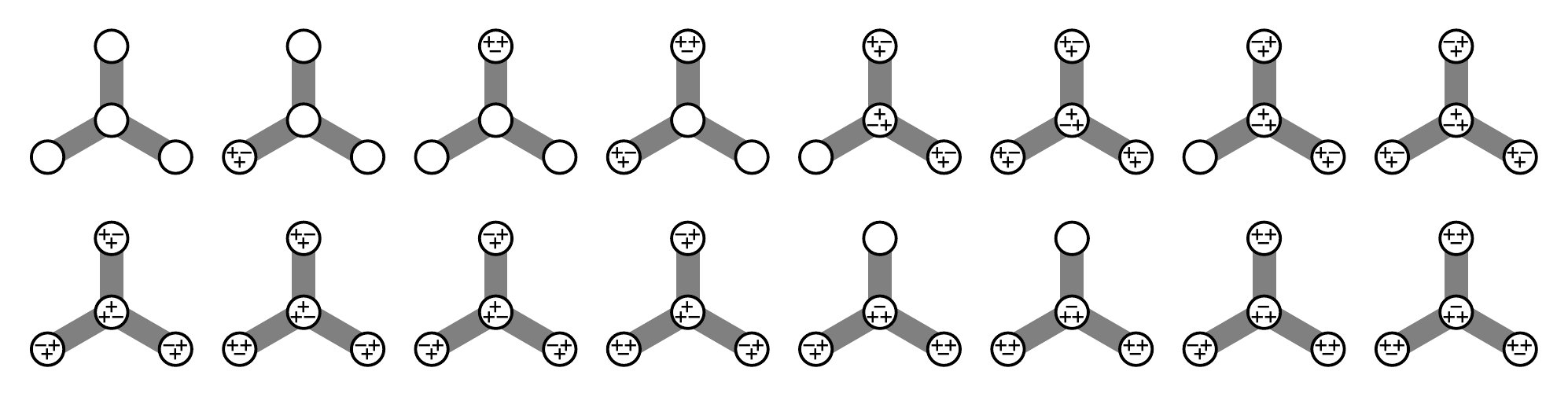}
\par\end{centering}
\caption{\label{fig:projected} States $\lt|\psi\lt(i,i_{x},i_{y},i_{z}\rt)\rira $
with $i$ on the even sublattice that satisfy $P^{\prime}_{i}\lt|\psi\lt(i,i_{x},i_{y},i_{z}\rt)\rira =\lt|\psi\lt(i,i_{x},i_{y},i_{z}\rt)\rira $.
For all other four-site states $\lt|\phi\rira $, $P^{\prime}_{i}\lt|\phi\rira =0$.
Projected states for the stars centered on odd sublattices are obtained by inversion. Empty circles represent $\lt|a^{x},a^{y},a^{z}\rira =\lt|-,-,-\rira $.}
\end{figure*}

Any perturbation that commutes with string operators such as those
in Eq. (\ref{eq:str_gs}) but not with the plaquette operators $W_{p}^{\s}$
lifts the ground state degeneracy.
As a warm-up example, we consider the longitudinal field
\begin{equation}
H_{\text{ext}}=h\sum_{i}A_{i}^{z}.
\label{eq:Hext}
\end{equation}
This perturbation quantifies laser tuning effects that localize Rydberg
excitations along predefined lines, an experimentally feasible setup
\citep{LabuhnPRA2014}. Such a field commutes with $V_{l_{xy}}^{\s}$
but not with $W_{p}^{\s}$ or $V_{l_{yz}}^{\s}$ and $V_{l_{xz}}^{\s}$, immediately implying that the model
$H_{\text{SOKM}}+H_{\text{ext}}$ has its degeneracy lifted to $2^{L_{xy}}$.
Excitations can be created on $x$ and $y$ bonds
 [see Fig. \ref{fig:AFM_figures}(d)] and then separated
without energy penalty if they are kept along the $xy$ chain. The
same excitations do not propagate along the transverse direction,
implying that the model now hosts excitations with restricted mobility
(or lineons). However, $H_{\text{ext}}$ does not induce immobile
excitations, which is thus insufficient to introduce fracton physics.

\begin{figure}
\begin{centering}
\includegraphics[width=0.5\columnwidth]{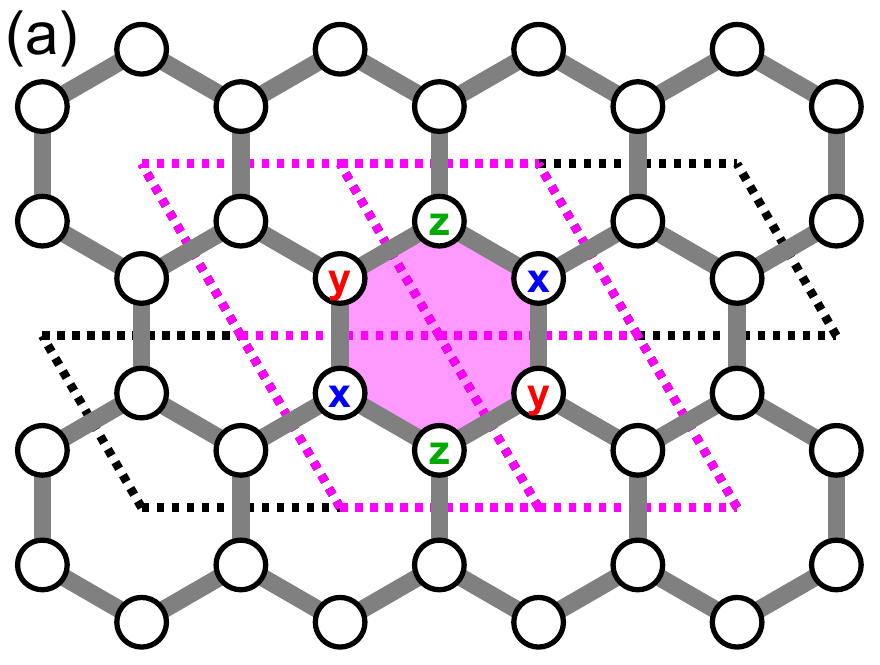}\includegraphics[width=0.5\columnwidth]{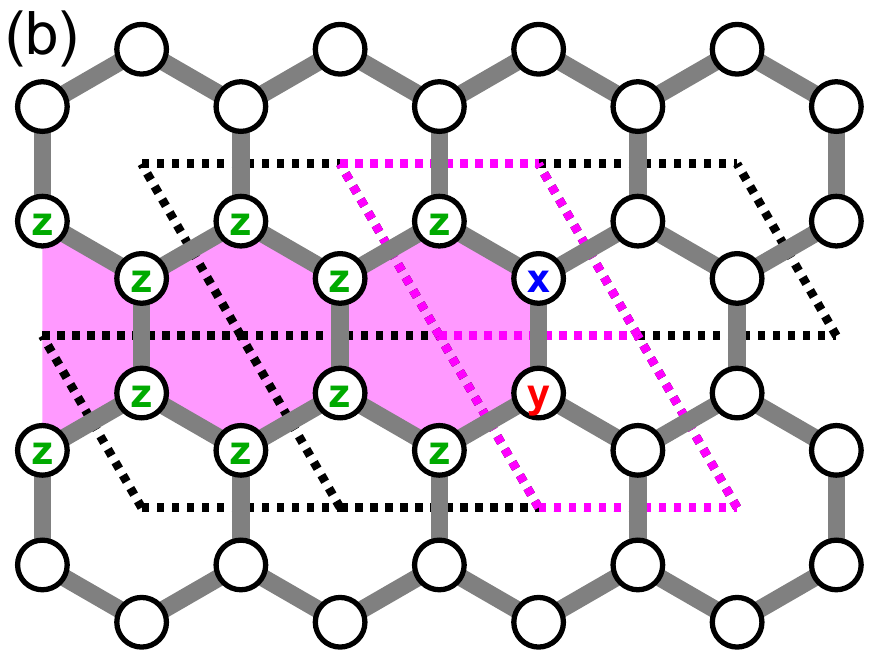}
\par\end{centering}
\begin{centering}
\includegraphics[width=0.5\columnwidth]{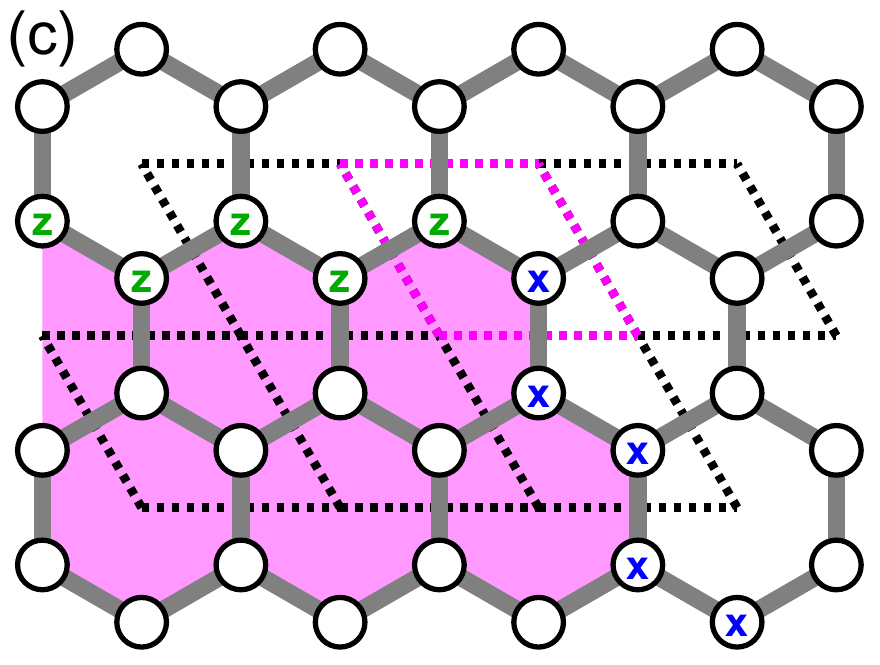}\includegraphics[width=0.5\columnwidth]{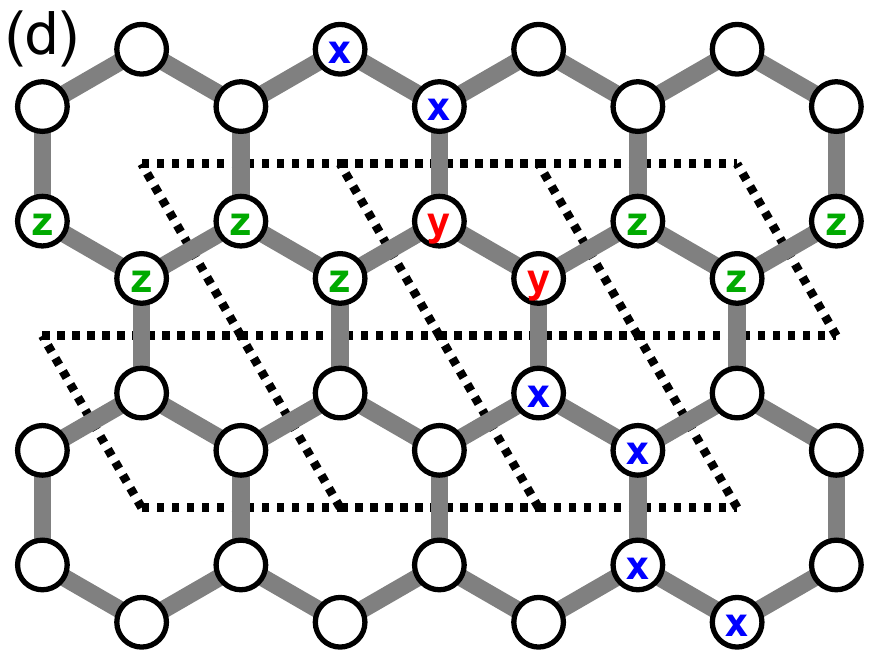}
\par\end{centering}
\begin{centering}
\includegraphics[width=0.8\columnwidth]{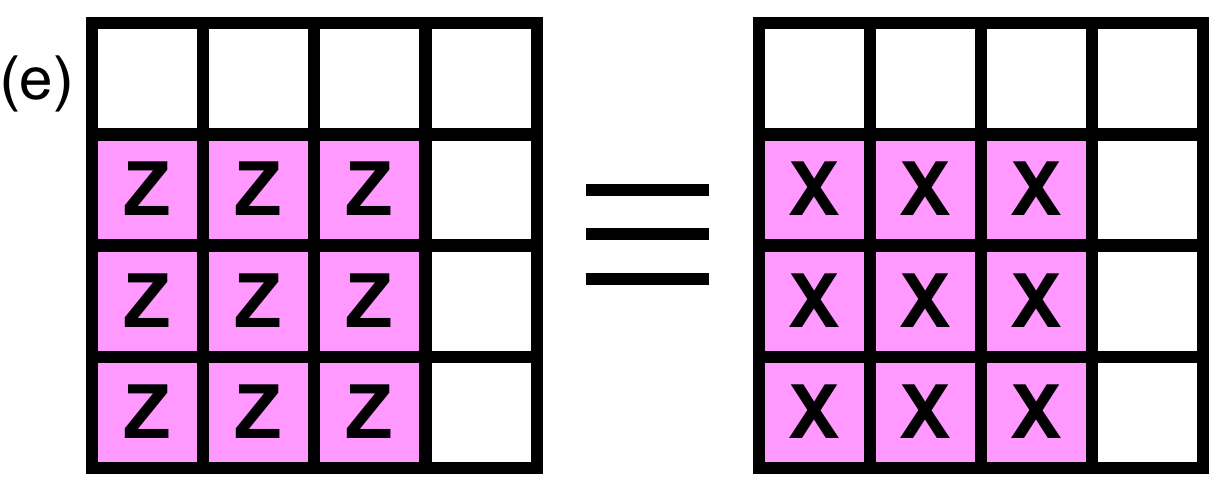}
\par\end{centering}
\caption{Excited states of the fracton model in Eq. (\ref{eq:fracton_H}) on the zero-flux sector. Magenta hexagons represent plaquettes where the $W_p^{\s}$ operator was applied in relation to the reference ground state.  (a) The excited state generated by a single $W_p^{\s}$. In the dual-lattice description, it frustrates the four rhomboids touching the central plaquette. (b) Lineon excitation generated from the application of $W_p^{\s}$ over a line of plaquettes leading to two unsatisfied rhomboids. (c) Classical fracton with a single unsatisfied rhomboid. (d) Effect of the four-dimensional local Hilbert space in terms of string operator intersections. (e) Representation of the classical fracton on the dual lattice. We emphasize that a given state in this dual representation can be represented in different ways. \label{fig:immobile} }
\end{figure}

Immobile excitations arise once we introduce a perturbation commuting
with strings in \emph{two} different directions. To construct an operator
with the desired property, consider the $\lt|\phi_{0}\rira $
state as well as other states obtained by the action of $V_{l_{xy}}^{\s},V_{l_{yz}}^{\s}$
strings. In all these states, the four-site stars centered at a given site $i$ have 16 different possibilities, all of which are indicated
in Fig. \ref{fig:projected} (apart from the inversion symmetry leading
to four-site states centered on the other sublattice). Let $P_{i}^{\prime}$
be the four-site projection operator satisfying $P_{i}^{\prime}\lt|\xi_{i}\rira =\lt|\xi_{i}\rira $,
if $\lt|\xi_{i}\rira $ is one of the states in Fig. \ref{fig:projected},
and $P_{i}^{\prime}\lt|\xi_{i}\rira =0$ otherwise. In terms
of $A_{i}^{\al}$ operators, we have
\begin{align}
P_{i}^{\prime} & =\frac{1}{16}+\frac{1}{16}\lt(A_{i}^{x}A_{i_{x}}^{x}+A_{i}^{y}A_{i_{y}}^{y}+A_{i}^{z}A_{i_{z}}^{z}\rt)\nonumber \\
 & +\frac{1}{16}\lt(A_{i}^{x}A_{i_{y}}^{x}+A_{i_{x}}^{x}A_{i_{y}}^{x}+A_{i}^{z}A_{i_{y}}^{z}+A_{i_{y}}^{z}A_{i_{z}}^{z}\rt)\nonumber \\
 & -\frac{1}{16}\lt(A_{i}^{x}A_{i_{y}}^{y}A_{i_{z}}^{z}+A_{i_{y}}^{x}A_{i}^{y}A_{i_{z}}^{z}+A_{i_{x}}^{x}A_{i}^{y}A_{i_{y}}^{z}+\rt.\nonumber \\
 & \lt.+A_{i_{x}}^{x}A_{i_{y}}^{y}A_{i}^{z}+A_{i_{x}}^{x}A_{i}^{y}A_{i_{z}}^{z}+A_{i_{x}}^{x}A_{i_{y}}^{y}A_{i_{z}}^{z}\rt)\nonumber \\
 & +\frac{1}{16}\lt(A_{i}^{x}A_{i_{x}}^{x}A_{i_{y}}^{z}A_{i_{z}}^{z}+A_{i_{x}}^{x}A_{i_{y}}^{x}A_{i}^{z}A_{i_{z}}^{z}\rt),\label{eq:Pprime}
\end{align}
in which $i$ is the site at the center of the star and $i_{\g}$
is the neighboring site along the $\g$ bond. Such a projector
commutes with $V_{l_{xy}}^{\s}$ and $V_{l_{yz}}^{\s}$ by construction.
 It is also clear from  the first line of Eq. (\ref{eq:Pprime}) that
some of its terms   commute with all $W_{p}^{\s}$. Let us
then consider two neighboring stars, one centered on the even and
other on the odd sublattice, as  depicted in Fig. \ref{fig:equivalences}(b).
The interactions in $P_{i}^{\prime}$ that simultaneously anticommute
with $W_p^\s$ for all four plaquettes shown in Fig.  \ref{fig:equivalences}(b)  are
\begin{align}
P_{i} & =\frac{1}{16}\lt(A_{i}^{x}A_{i_{y}}^{x}+A_{i_{x}}^{x}A_{i_{y}}^{x}+A_{i}^{z}A_{i_{y}}^{z}+A_{i_{y}}^{z}A_{i_{z}}^{z}\rt)\nonumber \\
 & -\frac{1}{16}\lt(A_{i_{y}}^{x}A_{i}^{y}A_{i_{z}}^{z}+A_{i_{x}}^{x}A_{i}^{y}A_{i_{y}}^{z}\rt)\nonumber \\
 & +\frac{1}{16}\lt(A_{i}^{x}A_{i_{x}}^{x}A_{i_{y}}^{z}A_{i_{z}}^{z}+A_{i_{x}}^{x}A_{i_{y}}^{x}A_{i}^{z}A_{i_{z}}^{z}\rt).\label{perturb}
\end{align}
We then define the classical fracton model  as follows:
\begin{equation}
H=-J\sum_{\lela ij\rira _{\g}}A_{i}^{\g}A_{j}^{\g}-\lambda\sum_{i}P_{i}.\label{eq:fracton_H}
\end{equation}

A simplified minimal model can be obtained in the zero-flux sector considering the dual lattice formed by the centers of the hexagons. Let the dual-lattice states be distinguished by how many times the $W_p^{\s}$ operators are applied over the reference state $\lt|\phi_{0}\rira $. Inserting one $W_{p}^{\s}$ as illustrated
in Fig. \ref{fig:immobile}(a) generates excitations on eight stars
centered on the sites located inside the highlighted rhomboids. We
can reduce the number of excitations by four stars if a semi-infinite
line of excitations is inserted, as indicated in Fig. \ref{fig:immobile}(b). Note that the pair of $x,y$ states at the end of the line in   Fig. \ref{fig:immobile}(b) can move as a lineon. 
The minimum number of excitations in the vicinity of the original plaquette is obtained after applying $W_p^\s$ as illustrated in Fig. \ref{fig:immobile}(c). This operation defines extended membrane operators, whose edges run along the directions of $xy$ and $yz$ chains.  The isolated excitations live at the corners of the membrane and cannot be moved without creating additional excitations. This behavior is captured by the effective model
\begin{equation}
H_{\text{eff}}=-J\sum_{\langle ijkl\rangle}\mc{Y}_{i}\mc{Y}_{j}\mc{Y}_{k}\mc{Y}_{l},\label{eq:Heff}
\end{equation}
in which $i,j,k,l$ label positions on the dual  lattice  corresponding to the centers of four adjacent hexagons that form a rhombohedron; see the four plaquettes indicated in Fig. \ref{fig:equivalences}(b) or the dual lattice drawn in Fig. \ref{fig:immobile}(a-d). The Ising variable $\mc{Y}$ is defined over the ferromagnetic reference ground state, being $\mc{Y}_{i}=-1$
if a plaquette operator was inserted; otherwise, $\mc{Y}_{i}=1$.

The Hamiltonian in Eq. (\ref{eq:Heff}) is remarkably similar to the
square-lattice model studied in Ref. \citep{HanYan2019},
in the sense that the interaction is a product of four Ising variables
defined at the center of four plaquettes. In fact, it can be understood as an SSSB phase, similar to that of the Xu-Moore model \cite{Xu2004, Xu2005} (or plaquette Ising model \cite{You2018}). A key difference concerns
the ground state degeneracy, which is equivalent to $2^{L_{xy}+L_{yz}-1}$
in Ref. \citep{HanYan2019}, whereas it is $2^{L_{xy}+L_{yz}}$ in
the present model. The root of this difference lies in the four-dimensional
Hilbert space at each site, which implies that at the intersection
of $V_{l_{xy}}^{\s}$ and $V_{l_{yz}}^{\s}$ we do not find the
$\lt| -,-,-\rira $ state, but rather the $\left|y\right\rangle$ one; see Fig. \ref{fig:immobile}(d).
We can account for this difference by treating $\mc{Y}_{i}$
as a four-state operator just like $A_{i}^{\g}$. Using the ordered
basis $\lt\{ \lt|0\rira ,\lt|X\rira ,\lt|Y\rira ,\lt|Z\rira \rt\} $,
we write
\begin{equation}
\mc{Y}=\text{diag}\lt(+,-,+,-\rt).\label{mcY}
\end{equation}
The $\lt|0\rira $ state corresponds to the absence of $W_{p}^{\s}$,
while an isolated $\lt|X\rira $ ($\lt|Z\rira $)
state corresponds to the combination of two lineons limited by $yz$
($xy$) chains. The $\lt|Y\rira $ state corresponds to
the absence of $W_{p}^{\s}$ due to the intersection between $\lt|X\rira $
and $\lt|Z\rira $ at a given point of the dual lattice.
All zero-flux states satisfying the bond energy can be mapped onto
a state on the dual lattice, as exemplified by the fracton state in
Fig. \ref{fig:immobile}(e). 

\section{Effective field theory\label{continummtheory}}

The study of fractons initially emerged through exactly solvable lattice spin models \cite{ChamonPRL2005,Vijay2015}, but has since been extended to include effective field theory approaches \cite{NandkishoreHermele2019}. One motivation for this type of approach is to shed light on robust, universal properties of the fractonic phase of matter that persist beyond the  solvable models. Unlike topological quantum field theories that describe topologically ordered states such as gapped QSLs, effective field theories for fractonic systems must retain certain geometric features and discrete symmetries of the lattice; see, e.g. Ref. \cite{Slagle2017}. In fact, physical properties associated with the low-energy sector, such as the ground state degeneracy, can depend on microscopic details related to the lattice regularization. Although the continuum limit may be more subtle in these cases \cite{seiberg2020,Seiberg2021,Seiberg2021a,Gorantla2023}, it is possible to derive an effective field theory starting from the lattice model by representing the physical spin operators as exponentials of fields that obey canonical commutation relations \cite{Fontana2021}. This representation naturally introduces an emergent gauge structure. From the perspective of the gauge theory, the ground state degeneracy and mobility constraints of excitations in the fractonic phase follow from subsystem conservation laws that can be inferred by analyzing the gauge invariance of the effective action. In particular, rigid Wilson lines constructed from the gauge fields play a central role in encoding these properties.

In this section, we construct and analyze an effective field theory that captures the main properties of the lattice model discussed in Secs. \ref{sec:The-Classical-Spin-Orbital} and \ref{Sec:Classical fractons}, notably  the existence of conserved line operators and the associated ground state degeneracy.

We start by  rewriting the SOKM Hamiltonian  as \begin{equation}
H_{\rm SOKM}=-J\sum_{\gamma}\sum_{\mb R_\gamma} O({\mb R_\gamma}),
\end{equation}
where $\mb R_\gamma$ are the positions of the centers of the $\gamma$ bonds and $O(\mb R_\gamma)=A^\gamma(\mb R_\gamma+\frac12\boldsymbol\delta_\gamma)A^\gamma(\mb R_\gamma-\frac12\boldsymbol\delta_\gamma)$.  Here we use the notation $A_j^\gamma=A^\gamma(\mb r_j)$ for the operators that act on site $j$, at  position $\mb r_j$ in the honeycomb lattice.  For fixed $\gamma$, the set  $\{\mb R_{\gamma}\}$ forms a triangular lattice. Crucially, $O(\mb R_\gamma)$ are mutually commuting operators that square to the identity. Thus, the SOKM belongs to the class of frustration-free models whose ground state subspace is a stabilizer code. This property is required to apply the procedure described in Refs.  \cite{Fontana2021, Fontana2023}, as we shall do in the following.  The first step is to   introduce a new basis of operators:  
\begin{equation}
 \Gamma^1=\sigma^x \tau^x,\; \Gamma^2=\sigma^y \tau^x,\; \Gamma^3=\sigma^z \tau^x, \; \Gamma^4=\sigma^0 \tau^y,
	\label{principalbasis}
\end{equation}
where $\sigma^0$ denotes the identity matrix in pseudospin space. We also define  $\Gamma^5=\Gamma^1\Gamma^2\Gamma^3\Gamma^4$. These operators  obey $\{\Gamma^I,\Gamma^J\}=2\delta_{IJ}$ for $I,J\in \{1,\dots,5\}$. The   operators $A^\gamma$ can be written as  \begin{equation}
A^x=\Gamma^1,\quad A^y=i\Gamma^1\Gamma^3\Gamma^4,\quad A^z=-i\Gamma^3\Gamma^4.
\end{equation}

Next, we represent the operators $\Gamma^I$  using four bosonic fields.  At each site,  we introduce the  fields $\theta_n(\mb r_j)$, with $n\in\{1,\dots,4\}$, which obey the commutation relation \begin{equation}
[ \theta_{m}(\mb r), \theta_{n}(\mb r')]=i\pi \delta_{\mb r\mb r'}(K^{-1})_{mn},\label{bosonalgebra}
\end{equation}
where $K$ is the    antisymmetric matrix\begin{equation}
K=\left(\begin{array}{cccc}
0&1&1&1\\
-1&0&1&1\\
-1&-1&0&1\\
-1&-1&-1&0
\end{array}\right).
\end{equation}
Following Refs. \cite{Fontana2021, Fontana2023}, we use the ``bosonization'' map\begin{equation}
\Gamma^I_{j} \sim \exp[it^I_mK_{mn}\theta_n(\mb r_j)],\label{boson}
\end{equation}
with an implicit sum over repeated indices $m,n\in \{1,\dots,4\}$. The vectors $t^I_m$  are chosen so that  \begin{align}
t^I_mK_{mn}t^J_n=\begin{cases}
		0 \text{ (mod 2)}, ~~~\text{if}~~~I=J,\\
		 1 \text{ (mod 2)}, ~~~\text{if}~~~I\neq J,
	\end{cases}
	\label{cond1}
\end{align}
and $\sum_{I=1}^5t_m^I=0$. A convenient  choice is \begin{eqnarray}
t_m^I&=&\delta_{Im}, \quad \text{for } I=1,\dots,3,\nonumber\\
t^4_m&=&-\delta_{m4}.\label{tI}
\end{eqnarray}
We write the local operators of the SOKM as  \begin{equation}
A^\gamma(\mb r_j)\sim \cos[\theta^\gamma(\mb r_j)],\label{Acos}
\end{equation}
where $\theta^\gamma(\mb r)=v_m^\gamma K_{mn} \theta_n(\mb r)$, with \begin{eqnarray}
v_m^x&=&t_m^1,\nonumber\\
v_m^y&=&-t_m^1+t_m^3+t_m^4,\nonumber\\
v_m^z&=&-t_m^3-t_m^4.\label{tgamma}
\end{eqnarray}
The four-component vectors  $v_n^\gamma$ are linearly dependent and satisfy   \begin{equation}
v^\gamma_nK_{mn}v^{\gamma'}_n=0.\label{vKiv} 
\end{equation} 
We then have $[\theta^\gamma(\mb r),\theta^{\gamma'}(\mb r')]=0$. This commutation relation  ensures that the operators $A^\gamma(\mb r)$ defined in Eq. (\ref{Acos}) commute among themselves, as happens in the original lattice model.

In the bosonic representation, the    SOKM  becomes\begin{eqnarray}
H_{\rm SOKM}&=& -\frac{J}2\sum_{\gamma}\sum_{\mb R_\gamma}\cos[2\theta^\gamma(\mb R_\gamma)]\nonumber\\
&& -\frac{J}2\sum_{\gamma}\sum_{\mb R_\gamma}  \cos[\Delta_\gamma\theta^\gamma(\mb R_\gamma)],\label{Heff1}
\end{eqnarray}
where \begin{eqnarray}
\theta^\gamma(\mb R_\gamma)&=&\frac{\theta^\gamma_{\rm e}(\mb R_\gamma+\boldsymbol \delta_\gamma/2)+\theta^\gamma_{\rm o}(\mb R_\gamma-\boldsymbol \delta_\gamma/2)}2,\\
\Delta_\gamma\theta^\gamma(\mb R_\gamma)&=&\theta^\gamma_{\rm e}(\mb R_\gamma+\boldsymbol \delta_\gamma/2)-\theta^\gamma_{\rm o}(\mb R_\gamma-\boldsymbol \delta_\gamma/2).
\end{eqnarray}
Here we have added the subindex e,o to indicate that the that fields act on positions that belong to the even or odd sublattices of the honeycomb lattice, respectively. At this point, we assume that we can take the continuum limit with   fields that vary smoothly on the lattice scale. Similar assumptions have been made in the derivation of effective field theories for fractonic models such as the X-cube \cite{Slagle2017} and the Chamon code \cite{Fontana2021}. Although our system has ultrashort-range correlations---as usual for stabilizer codes---this continuum limit can be justified a posteriori by showing that the resulting field theory reproduces key properties of the solvable lattice model. In this approximation, we replace  $\Delta_\gamma\theta^\gamma \to \frac1{\sqrt3}a_0\partial_\gamma \theta^\gamma(\mb r)$, where $a_0$ is the lattice spacing (set to $a_0=1$ in previous sections) and  $\partial_\gamma=\hat{\boldsymbol \delta}_\gamma\cdot \nabla$ are spatial   derivatives in the direction of the unit vectors $\hat{\boldsymbol \delta}_\gamma=\sqrt3 \boldsymbol \delta_\gamma/a_0$. Expanding the second term in Eq. (\ref{Heff1}) to leading order in the derivatives, we  obtain the effective Hamiltonian in the continuum limit 
\begin{equation}
H_{\rm c}= \sum_\gamma \int d^2r\, \left[\frac{u}{2}(\partial_\gamma \theta^\gamma)^2-g \cos(2\theta^\gamma)\right],\label{Heff2}
\end{equation}
where $u\sim J$ and $g\sim J/a_0^2$. This Hamiltonian  respects the $Z_3$ symmetry that takes $\gamma\mapsto \gamma+1$ (mod 3)  in real and internal space, thus retaining the discrete rotational symmetry of the microscopic model.  The cosine term imposes the constraint that the physical values of the phase fields are $\theta^\gamma=0,\pi$ (mod $2\pi$), corresponding to the eigenvalues $a^\gamma=\pm1$ of $A^\gamma$. If this term is irrelevant, we can treat $\theta^\gamma$ as continuous functions and focus on the effects of the first term in Eq. (\ref{Heff2}). 

We can reformulate the effective field theory as a gauge theory.  Let us consider the effective action\begin{equation}
S=\int_{\mc M} d^3x\, \left(\frac{K_{mn}}{4\pi} \mc A_m  \partial_t\mc A_n -\frac1{2\pi}\sum_\gamma  \mc A_0^\gamma \partial_\gamma \mc A^\gamma\right), \label{action}
\end{equation}
where $\mc M$ is a spacetime manifold,  $\mc A_n(\mb r)=\theta_n(\mb r)/a_0$ are rescaled fields with dimension of inverse length, with the linear combinations $\mc A^\gamma (\mb r) =v_m^\gamma K_{mn}\mc A_n(\mb r)$, and $\mc A_0^\gamma$ are Lagrange multipliers enforcing the constraint $\partial_\gamma \mc A^\gamma=0$ in the ground state subspace.  This action is reminiscent of the Chern-Simons-like theories studied in Refs. \cite{Fontana2021, Fontana2022, Fontana2023, Delfino2023}. In fact, using Eq. (\ref{vKiv}) we  verify that the action is invariant under the U(1) gauge transformation \begin{eqnarray}
\mc A_m&\mapsto& \mc A_m + \sum_\gamma v_m^\gamma\, \partial_\gamma G^\gamma ,\\
\mc A_0^\gamma&\mapsto& \mc A_0^\gamma  + \partial_tG^\gamma ,
\end{eqnarray}
where $G^\gamma(\mb r,t)$ are arbitrary dimensionless functions. Explicitly, the gauge transformation acts on the fields as \begin{eqnarray}
\mc A_1&\mapsto&  \mc A_1 +\partial_1G^x -\partial_2G^y,\\
\mc A_2&\mapsto&  \mc A_2,\\
\mc A_3&\mapsto & \mc A_3 +\partial_2G^y -\partial_3G^z,\\
\mc A_4&\mapsto&  \mc A_4 -\partial_2G^y +\partial_3G^z.
\end{eqnarray} 
Defining the linear combinations $\mc A_\pm=(\mc A_3\pm \mc A_4)/2$, we note that $\mc A_2$ and $\mc A_+$ are gauge invariant. In order for $e^{i\mc A_n(\mb r)}$ to describe physical operators for all components $\mc A_n$, the gauge functions must be restricted to the form \begin{equation}
G^\gamma(\mb r)=2\pi n^\gamma(\bar x_\gamma)\,x_\gamma +f^{\gamma}(\bar x_\gamma),
\end{equation}
where $ x_\gamma=\mb r\cdot \hat{\boldsymbol\delta}_\gamma$ is the coordinate in the direction parallel to  the $\gamma$ bond, $\bar x_\gamma$ is the coordinate in the perpendicular direction, $n^\gamma(\bar x_\gamma)$ are integer functions, and $f^\gamma(\bar x_\gamma)$ are arbitrary real functions.  This type of behavior   is common in higher-rank gauge theories, where it stems from  the structure of higher derivatives and allows for discontinuous configurations in the low-energy sector  \cite{Seiberg2021, Seiberg2021a, Rudelius2021, Gorantla2023}.

The equations of motion derived from the action in Eq. (\ref{action}) lead to the conditions of vanishing   ``electric'' and ``magnetic'' fields: \begin{eqnarray}
\mc E_{n}&\equiv &\partial_t \mc A_n -\sum_\gamma v_n^\gamma \partial_\gamma \mc A_0^\gamma=0,\label{electric}\\
\mc B^\gamma&\equiv & K_{mn}D^\gamma_m  \mc A_n=0,
\end{eqnarray}
where $ D^\gamma_m=v_m^\gamma \partial_\gamma$.  From Eq. (\ref{electric}), we obtain \begin{eqnarray}
\partial_t\mc A_2&=&\partial_t\mc A_+=0,\nonumber\\
\partial_t\mc A_1&=&\partial_1 \mc A_0^x-\partial_2 \mc A_0^y,\nonumber\\
\partial_t\mc A_-&=&\partial_2 \mc A_0^y-\partial_3 \mc A_0^z.
\end{eqnarray}
The conservation of $\mc A_2(\mb r)$ and $\mc A_+(\mb r)$ has a simple physical interpretation. In this classical theory, the Hamiltonian is a function of the mutually commuting fields $\theta^\gamma(\mb r)$. These fields are linearly dependent and  can be written in terms of only two independent components.  For the parametrization in Eqs. (\ref{tI}) and (\ref{tgamma}), we have $\theta^x=\theta_2+2\theta_+$, $\theta^y=-\theta_2$, and $\theta^z=-2\theta_+$. Thus, the conservation of $\theta^\gamma$ implies  that $\mc A_2$ and $\mc A_+$  become conserved quantities in the gauge theory.

 The theory also contains local operators that involve the  fields $\theta_1$ and $\theta_-=(\theta_3-\theta_4)/2$ and act nontrivially on the eigenvalues of  $A^\gamma\sim \cos(\theta^\gamma)$.   For instance, we have\begin{equation}
 \sigma_j^z=-i\Gamma^1_j\Gamma^2_j\sim e^{-i\theta_1},
 \end{equation}
 where we omit the factors that depend on   $\theta_2$ and $\theta_+$, which only give unimportant phases when acting on a ground  state. The operator $e^{-i\theta_1}$ anticommutes with $\cos(\theta^{x,y})$ on the same site, but commutes with $\cos(\theta^z)$. We can also check that $e^{-i\theta_1}$ commutes with the terms  $\cos(2\theta^\gamma)$  in the effective Hamiltonian in Eq. (\ref{Heff2}). Similarly,  $\sigma_j^x\sim e^{-i(2\theta_1+\theta_-)}$ anticommutes  with $\cos(\theta^{y,z})$, but commutes with $\cos(\theta^x)$. Once again, these relations confirm that the mapping in Eq. (\ref{Acos}) reproduces the behavior observed in the lattice model.
 
 We can now identify the line operators in the effective field theory. Consider the model defined on a torus with $L_{yz}$ unit cells  in the direction of $\hat{\mb e}_1=\hat{\mb x}$ and $L_{xy}$ unit cells in the direction of $\hat{\mb e}_2=-\frac{1}2\hat{\mb x}+\frac{\sqrt 3}{2}\hat{\mb y}$. Note that $\mb r\cdot \hat{\mb e}_1=\bar x_3$ and $\mb r\cdot \hat{\mb e}_2=\bar x_1$. The line operator associated with applying $\sigma_j^z$ along an $xy$ chain that winds around the torus has the form\begin{equation}
  V^\sigma_{l_{xy}}\sim \prod_{k=0}^{L_{yz}-1} e^{-i\theta_1(\mb r+ka_0\hat{\mb e}_1)}e^{-i\theta_1(\mb r+\boldsymbol\delta_1+ka_0\hat{\mb e}_1)},
 \end{equation}
 where $\mb r$ belongs to the odd sublattice.  Note that each line in the direction of $\hat{\mb e}_1$ has length $\ell_1=L_{yz}a_0$ and  can be labeled by $(\mb r+ka_0\hat{\mb e}_1)\cdot \hat{\boldsymbol \delta}_3=x_3$.   Thus, in the continuum limit, \begin{equation}
  V^\sigma_{l_{xy}}\to \mc V_{z}(x_3)\sim e^{-i\oint d\bar {x}_3 (\mc A_{1}^{\rm o}+\mc A_{1}^{\rm e})}.
 \end{equation}
 Here we have restored the indices $\nu =\text{e},\text{o}$ for even and odd sublattices, with the convention $\mc A_n=(\mc A_n^{\rm o}+\mc A_n^{\rm e})/2$.  Similarly, the line operator acting on a $yz$ line with length $\ell_2=L_{xy}a_0$ is given in the continuum limit  by 
 \begin{equation}
   V^\sigma_{l_{yz}}\to\mc V_{x}(x_1) \sim  e^{-i\oint d\bar x_1 (2\mc A_{1}^{\rm o}+2\mc A_{1}^{\rm e}+\mc A_{-}^{\rm o}+\mc A_{-}^{\rm e}) }.
 \end{equation}
Using the commutation relation  \begin{equation}
[\mc A^\nu_{m}(\mb r),\mc A_{n}^{\nu'}(\mb r')]=i\delta_{\nu\nu'}(K^{-1})_{mn}\delta(\bar x_1-\bar  x_1')\delta(\bar x_3-\bar  x_3'),
\end{equation}
we verify that 
\begin{equation}
 [ \mc V_{z}(x_3), \mc V_{x}(x_1)]=0,\qquad \forall\, x_1,x_3. 
 \end{equation}
Importantly,  line operators running in different directions commute with each other because their crossing contains two sites, one in each sublattice.  To see that they also commute with the Hamiltonian, we need to invoke the lattice regularization and interpret the spatial derivative in Eq. (\ref{Heff2}) as $\partial_\gamma\mc A^\gamma=\sqrt3( \mc A^{\gamma,\rm e}-\mc A^{\gamma,\rm o})/a_0$, noting  that conjugation by a line operator shifts $\mc A^{\gamma,\nu}$ by the same amount for both sublattices. 

In addition to  the line operators $V_l^\sigma$, the SOKM   has conserved quantities associated with contractible loops because the lines along which we apply the $\sigma_j^\gamma$ operators can make $120^\circ$ turns  without creating any excitations. 
The perturbation in Eq. (\ref{perturb}) breaks the conservation of the line operators $ V^\sigma_{l_{xz}}$ and associates an energy cost with the corners between the $l_{xy}$ and $l_{yz}$ lines. As a result, contractible loops are excluded from the ground state sector, and the ground states are generated by rigid line operators acting upon the reference state with uniform $A_j^\gamma$.   From the perspective of the effective field theory, the perturbation generates terms that break the $Z_3$ symmetry but still commute with $\mc V_{z}(x_3)$   and $\mc V_{x}(x_1)$.

The ground state degeneracy is associated with the nontrivial action of the line operators on  the local conserved quantities of the classical model. For instance,\begin{equation}
\mc V_{z}(x_3)e^{ia_0\mc A^{\gamma}(\mb r')}=e^{i\pi a_0\delta(x_3-x_3')}e^{ia_0\mc A^{\gamma}(\mb r')}\mc V_{z}(x_3).\label{linepoint}
\end{equation} 
Note that, when written in terms of $\mc A_n(\mb r)$, the local conserved quantities involve the short-distance scale $a_0$, and the correct anticommutation relation requires the regularization $a_0\delta(x_3-x_3')\to 1$ for $x_3\to x_3'$, i.e., when the line operator $\mc V_{xy}(x_3)$ crosses the point $\mb r'$. It is convenient to use the local conserved quantities to define another set of line operators:\begin{eqnarray}
\mc U_x(x_1)&=&e^{-i\oint d\bar x_1 \mc A^{x,\rm e}},\\
\mc U_z(x_3)&=&e^{-i\oint d\bar x_3 \mc A^{z,\rm e}},
\end{eqnarray}
which act on  only one sublattice. We then obtain the nontrivial algebra \begin{eqnarray}
\mc U_x(x_1)\mc V_z(x_3)=-\mc V_z(x_3)\mc U_x(x_1),\label{line1}\\
\mc U_z(x_3)\mc V_x(x_1)=-\mc V_x(x_1)\mc U_z(x_3).\label{line2}
\end{eqnarray}
Equations   (\ref{line1}) and  (\ref{line2}) imply that the ground state degeneracy scales exponentially with the total  number of lines, as $ 2^{L_{xy}+L_{yz}}$. As usual in effective field theories for fractons  \cite{seiberg2020, Seiberg2021,  Fontana2021, Fontana2022, Fontana2023, Pretko2017, Pretko2017a, Slagle2017,Slagle2019, Burnell2022, Rudelius2021},  this counting depends on a lattice regularization since the number of lines is given by the lengths of the cycles divided by  the short-distance scale, $L_{yz}=\ell_1/a_0$ and $L_{xy}=\ell_2/a_0$.   


\section{Conclusion and Outlook\label{sec:conclusion}}

We unveiled the spectrum, level degeneracy, and partition function
of the classical SOKM, which is the parent Hamiltonian describing
quantum simulations of Rydberg atoms placed on the ruby lattice \citep{Verresen2022}.
The mapping between $S=3/2$ honeycomb models and spin models on the
kagome lattice allows an understanding of the SOKM entropy and specific
heat in terms of the kagome spin ice. We developed a low-energy field theory
 using methods to study $\Gamma$-matrix fracton models. We identified the underlying
gauge transformations and showed that the continuum theory recovers the ground state degeneracy of the lattice model. In addition to providing exact results for the SOKM, our symmetry analysis guided the proposal of a perturbation that keeps the model classical but induces immobile fractonic excitations. We expect that our work will guide further research on
realistic implementations of FSLs and possible pathways to connect them
with QSLs. Another interesting line of inquiry concerns the adaptability
of other simulators for kagome spin ice phases \citep{Tanaka2006,Lopez_Bezanilla_2023,Zhang2012}
to simulate the fracton model discussed here.

Our work indicates that classical integrable spin models with stable
FSLs can be formulated from exactly solvable models with Majorana
QSL ground states. In quantum models, the extensive $Z_{2}$ symmetries
can be exactly mapped onto a static $Z_{2}$ gauge field, thus allowing
their formulation in terms of free fermions. The symmetries in the
classical counterparts are responsible for their exponentially large
ground state degeneracy; being selective on how to break such symmetries
allows us to introduce fractons. This procedure can be immediately
applied to investigate extensions of the SOKM on other three-coordinated
lattices that are known to host Kitaev spin liquids \citep{YaoKivelsonChiral2007,OBrien2016,Peri2020,Cassella2023}
to verify the effect of lattice topology on the fracton properties.
In particular, the SOKM is exactly solvable in three-coordinated hyperbolic
lattices \citep{Dussel2024,Lenggenhager2024}, so that we expect that
the corresponding fracton theory satisfies key properties of the AdS/CFT
correspondence \citep{HanYan2019}. The Rydberg atoms can be arranged  in
all these lattices using a recently formulated protocol with programmable
tweezer arrays \citep{JuliaFarre2024}. As a further step, just as
the KHM principles can be extended to formulate exactly solvable 
models with spin $S>1/2$ \citep{Nussinov2009,Chulliparambil2020,Chulliparambil2021},
the SOKM and its fracton model are expected to be extended to more
general $\Gamma$-matrix models.

The simplicity of the fracton model indicates that it can be efficiently simulated using Monte Carlo algorithms \citep{Placke2024}. Developing such
numerical methods are useful to study the SOKM with classical symmetry-breaking
 perturbations, most notably the longitudinal field $h_{l}$.
An interesting question concerns the possible induction of magnetization
plateaus through these fields akin to those observed in kagome
spin ice \citep{Moessner2003}. These numerical methods can also uncover
static and dynamic correlation functions, thus providing phase signatures
in a larger range of experimental parameters. 

Finally, we point out that the spin-orbital representation used throughout this paper can also aid the discussion of quantum fluctuations. The simplest example occurs for the case $\Omega = t^{\prime \prime}, t^{\prime} = 0$ in Eq. (\ref{eq:Hfl}), which was called $f$-anyon fluctuations in Ref. \cite{Verresen2022}. This term is translated into the spin-orbital notation by an out-of-plane external field $H_Z = -\Omega \sum_i(\tau^x_i+\tau^y_i+\tau^z_i)$. This result connects our study on quantum simulations with those on the Kitaev \cite{SunChen2009, SorensenPRR2023, SorensenIntegerSpin2023, Richards2024, Wang2025, Zhu_2025} and Yao-Lee \cite{Seifert2020,Chulliparambil2021} models under out-of-plane external fields. The rich phenomenology related to those quantum fluctuations will be the subject of future research.

\emph{Note} - As we concluded this manuscript, we came across an independent work studying the thermodynamics and static structure factor of the SOKM using the Monte Carlo method \citep{HanYan2025}. Our results are entirely consistent with their numerical simulations. 

\section*{Acknowledgments}

W. Natori gives a special thanks to Johannes Knolle for the first
studies on the SOKM. He also thanks Hui-Ke Jin for related work 
and Eric Andrade for useful discussions about the kagome spin ice. This research was supported by the MOST Young Scholar Fellowship
(Grants No. 112-2636-M-007-008- and No. 113-2636-M-007-002-), National
Center for Theoretical Sciences (Grants No. 113-2124-M-002-003-)
from the Ministry of Science and Technology (MOST), Taiwan, and the
Yushan Young Scholar Program (as Administrative Support Grant Fellow)
from the Ministry of Education, Taiwan. We also acknowledge support by the National Council for Scientific and Technological Development -- CNPq (R.G.P.) and by a grant from the Simons
Foundation (R.G.P., W.M.H.N.). Research at IIP-UFRN is supported by Brazilian ministries MEC and MCTI. 

\bibliographystyle{apsrev4-1_control}
\bibliography{Fractons}

\appendix

\section{Spin and Orbital Degrees of Freedom} \label{sec:app}

First, let us set the usual definition of the pseudospin and pseudo-orbital
degrees of freedom in $S=3/2$ systems following Ref. \citep{Natori2023}

\begin{align}
\lt(\sigma_{i}^{\al}\rt)^{\prime} & =-i\exp\lt(i\pi S_{i}^{\al}\rt),\nonumber \\
\lt(\tau_{i}^{z}\rt)^{\prime} & =\lt(S_{i}^{z}\rt)^{2}-\frac{5}{4},\nonumber \\
\lt(\tau_{i}^{x}\rt)^{\prime} & =\frac{1}{\sqrt{3}}\lt[\lt(S_{i}^{x}\rt)^{2}-\lt(S_{i}^{y}\rt)^{2}\rt],\nonumber \\
\lt(\tau_{i}^{y}\rt)^{\prime} & =\frac{2\sqrt{3}}{9}\overline{S_{i}^{x}S_{i}^{y}S_{i}^{z}}.
\end{align}
We considered above the canonical matrix representation of the $S=3/2$
spins. It is useful to rewrite this model in a different basis obtained
through the following unitary matrix 
\begin{equation}
U=\frac{1}{\sqrt{2}}\lt(\begin{array}{cccc}
e^{i\pi/4} & 0 & 0 & e^{-i\pi/4}\\
e^{i\pi/4} & 0 & 0 & -e^{-i\pi/4}\\
0 & e^{-i\pi/4} & e^{i\pi/4} & 0\\
0 & e^{-i\pi/4} & -e^{i\pi/4} & 0
\end{array}\rt),
\end{equation}
in which 
\begin{align}
U^{\dagger}\lt(\sigma_{i}^{x}\tau_{i}^{x}\rt)^{\prime}U=A^{x} & =\text{diag}\lt(-1,1,-1,1\rt),\nonumber \\
U^{\dagger}\lt(\sigma_{i}^{y}\tau_{i}^{y}\rt)^{\prime}U=A^{y} & =\text{diag}\lt(-1,1,1,-1\rt),\nonumber \\
U^{\dagger}\lt(\sigma_{i}^{z}\tau_{i}^{z}\rt)^{\prime}U=A^{z} & =\text{diag}\lt(-1,-1,1,1\rt).
\end{align}
Applying the unitary transformations $U^{\dagger}\lt(\sigma_{i}^{\g}\rt)^{\prime}U=\s^{\g}$,
$U^{\dagger}\lt(\tau_{i}^{\g}\rt)^{\prime}U=\tau^{\g}$ on
the pseudospin and pseudo-orbital representations yields 
\begin{align}
\s^{z}=\lt(\begin{array}{cccc}
0 & 0 & 0 & i\\
0 & 0 & -i & 0\\
0 & i & 0 & 0\\
-i & 0 & 0 & 0
\end{array}\rt), & \quad\s^{x}=\lt(\begin{array}{cccc}
0 & i & 0 & 0\\
-i & 0 & 0 & 0\\
0 & 0 & 0 & -i\\
0 & 0 & i & 0
\end{array}\rt),\nonumber \\
\s^{y}=\lt(\begin{array}{cccc}
0 & 0 & i & 0\\
0 & 0 & 0 & i\\
-i & 0 & 0 & 0\\
0 & -i & 0 & 0
\end{array}\rt),
\end{align}
\begin{align}
\tau^{z}=\lt(\begin{array}{cccc}
0 & 0 & 0 & -i\\
0 & 0 & -i & 0\\
0 & i & 0 & 0\\
i & 0 & 0 & 0
\end{array}\rt), & \quad\tau^{x}=\lt(\begin{array}{cccc}
0 & -i & 0 & 0\\
i & 0 & 0 & 0\\
0 & 0 & 0 & -i\\
0 & 0 & i & 0
\end{array}\rt),\nonumber \\
\tau^{y}=\lt(\begin{array}{cccc}
0 & 0 & -i & 0\\
0 & 0 & 0 & i\\
i & 0 & 0 & 0\\
0 & -i & 0 & 0
\end{array}\rt).
\end{align}

\end{document}